\documentclass[11pt]{article}
\usepackage[utf8]{inputenc}
\usepackage{changepage}
\usepackage{amsmath}
\usepackage{amssymb}
\usepackage[margin=1in, left=.5in, right=.5in]{geometry}
\usepackage{multirow}
\usepackage{commath}
\usepackage{enumitem}
\usepackage{graphicx}
\usepackage{titling}
\usepackage{float}
\usepackage[toc]{appendix}
\usepackage[super]{natbib}

\newcommand\independent{\protect\mathpalette{\protect\independenT}{\perp}}
\def\independenT#1#2{\mathrel{\rlap{$#1#2$}\mkern2mu{#1#2}}}

\begin{document}

\title{Addressing Positivity Violations in Causal Effect Estimation using Gaussian Process Priors}
\author{Yaqian Zhu, Nandita Mitra $^\spadesuit$, Jason Roy $^\spadesuit$}
\date{$^\spadesuit$Co-senior authors}

\maketitle

\abstract{In observational studies, causal inference relies on several key identifying assumptions. One  identifiability condition is the positivity assumption, which requires the probability of treatment be bounded away from 0 and 1. That is, for every covariate combination, it should be possible to observe both treated and control subjects, i.e., the covariate distributions should overlap between treatment arms. If the positivity assumption is violated, population-level causal inference necessarily involves some extrapolation. Ideally, a greater amount of uncertainty about the causal effect estimate should be reflected in such situations. With that goal in mind, we construct a Gaussian process model for estimating treatment effects in the presence of practical violations of positivity.  Advantages of our method include minimal distributional assumptions, a cohesive model for estimating treatment effects, and more  uncertainty associated with areas in the covariate space where there is less overlap. We assess the performance of our approach with respect to bias and efficiency using simulation studies. The method is then applied to a study of critically ill female patients to examine the effect of undergoing right heart catheterization.}

\section{Introduction}\label{sec1}

Researchers often aim to infer the causal effects of a treatment on a population of interest from observational studies. Identification of causal effects from observational data relies on assumptions including ignorability, often referred to as ``no unmeasured confounding," which holds when treatment assignment is random (that is, independent of potential outcomes) given measured confounders.\cite{hernan_2006} If the treatment received depends on observed covariates, then the distribution of these covariates is expected to differ by treatment group. This raises concerns about violations of a second identifying assumption called positivity. Positivity assumes that there is a non-zero probability of receiving  treatment for all individuals. If there is a subpopulation defined by covariates for which one of the treatments is not observed, causal contrasts for that subgroup cannot be identified without further assumptions.\cite{westreich_2010, imbens_2015}

Theoretical (or structural) violation of the positivity assumption occurs when a subpopulation of individuals have zero probability of receiving at least one of the treatments, so that even if we let the sample size go to infinity, we would still never observe all treatment values. This can happen, for example, when treatment is contraindicated in a certain subgroup of patients because of their age, comorbidities, and family history of disease.  D'Amour et al. (2020) elaborate on this type of violation in the context of high-dimensional covariates.\cite{damour_2020} On the other hand, practical (also called random) violation of positivity arises when, in a given observational data set, a subpopulation is not observed to receive a particular treatment by chance.  For example, suppose in a sample, no males between the ages of 35 to 45 receive treatment $A$ purely by chance. In reality, their probability of receiving treatment $A$ may be small but not zero. In this case, we will not be able to learn about the treatment effect of $A$ in this subpopulation of men without making additional modeling assumptions.  We expect practical positivity violations to arise in clinical data, especially when there are a large number of covariates.  However, we could potentially learn about these nonoverlap regions using modeling. For instance, if we are willing to assume an additive linear regression model, we could learn about males treated with treatment $A$ via linear interpolation or extrapolation from younger or older men who received treatment $A$. The disadvantage is that we would need to rely on strong parametric assumptions.\cite{king_2006} Further, these approaches  may underestimate the degree of uncertainty that would be expected in data-sparse regions. 

Trimming approaches are commonly used to address positivity violations and are discussed in Petersen et al. (2012).\cite{petersen_2012} Crump et al. (2009) propose a method that removes (trims)  subjects whose propensity scores are outside specified bounds and calculates a minimum variance estimate on the remaining subsample.\cite{crump_2009} This approach requires  correct specification of the propensity score model and may result in a final sample that is a small subsample of the original study population. Others (Rosenbaum (2012) and Visconti \& Zubizarreta (2018)) have suggested matching treated and control subjects on covariates or propensity scores;\cite{Rosenbaum_2012, visconti_2018} however, external validity may be diminished due to matching because the target population of interest will have changed to that of the matched population. Hill \& Su (2013) define nonoverlap as the set of subjects whose estimated individual causal effects have corresponding variances that are greater than specified upper bounds.\cite{hill_2013} However, these upper bound cut-offs may rely on user specifications and may not adequately reflect the amount of data sparsity. Ghosh (2018) characterizes multivariate covariate overlap using convex hulls to determine positivity violations.\cite{Ghosh_2018, GhoshCruzCortes_2019} This overlap subset is termed `the margin' and is determined using a propensity score model; subjects who are outside the margin are trimmed.  A disadvantage to these trimming procedures is that by discarding subjects, the target of inference may shift to a resulting  subpopulation which may not represent the original population of interest.

An alternative to removing subjects entirely is to downweight subjects in regions with less overlap. In that spirit, Li et al. (2018) proposed estimating causal effects using overlap weights. \cite{li_2018, li_2019} However, overlap weights involve propensity score estimation and place more emphasis on those with higher probabilities of receiving either treatment. Although trimming or weighting approaches may be appropriate for structural violations, they fall short for practical violations of the positivity assumption. In this practical violations setting, we expect covariates of subjects who violate this assumption to not be too far from the area of the covariate space where there is complete overlap. In other words, we consider the setting where our study objectives aim to make inference for a population but, by chance, the available sample of individuals does not include certain individual characteristics. For example, in studies that seek to inform public policy, population-level inference is desired since changes in policy will affect the general population. However, in a particular data set there could be some non-overlap by chance. Or, consider a study comparing the safety and effectiveness of drug A versus drug B. Suppose there happens to be no Hispanic individuals over age 60 who take drug A, but we expect that if we had a larger study that cell would no longer be empty. We do not want to exclude Hispanic individuals over age 60 from our analysis, but we do want our inferential procedures to account for the fact that contributions to the overall treatment effect estimate from this subpopulation will involve extra uncertainty due to extrapolation. In situations when there is an intended patient population for a treatment or intervention, but this population may not be entirely reflected in the available treatment group data, methods for population-level estimation are needed. The objective in these studies is often to obtain inference that preserves the original population, ensuring results may be generalized accordingly. When there are practical but not structural violations, interest generally centers on understanding the treatment effects for the entire population.  Approaches that account for positivity violations while also targeting a causal estimand for the whole population are, therefore, of most interest.

There has been some recent work on methods that are based on extrapolating information from overlap regions to nonoverlap regions in a way that preserves the intended target of inference. For instance, Nethery et al. (2019) propose a method for estimating a causal effect on the entire population using extrapolation.\cite{nethery_2019} Their definition of the overlap region relies on two user-specified parameters that indicate the desired extent of closeness between treatment groups based on subjects' estimated propensity scores. Choices regarding propensity score model specification and user inputs influence whether a subject is included in the overlap region. Having a fixed region means that uncertainty around subjects' inclusion in the overlap or nonoverlap region is ignored. In this approach, Bayesian additive regression tree (BART) models are used to estimate causal effects for the overlap regions and then in a subsequent stage, spline (SPL) models extrapolate those trends to subjects in the nonoverlap region. These choices in modeling mean that prior information on the treatment effect cannot be directly utilized. Lastly, although they account for the greater uncertainty in areas of nonoverlap, their proposed variance inflation strategy results in over-coverage as seen in their simulation studies. 

To address some of the above limitations of existing approaches, we propose a Gaussian process modeling approach for estimating average treatment effects in a way that preserves the original target of inference when there are practical positivity violations. \cite{rasmussen_2006, neal_1998} Our method contributes several advances to the current literature. First, because we use a non-parametric prior distribution, we avoid making parametric modeling assumptions.  Further, the prior does not rely on user-specified parameters nor cut-offs to define regions of overlap. The amount of nonoverlap is accounted for in the covariance functions of the Gaussian process as the distances of a subject's covariate values from those of individuals in the other treatment group. This provides us with a sense for nonoverlap that is data driven. We also avoid the need to construct explicit overlap and nonoverlap groups, allowing covariate distance and positivity violations to be considered in a more continuous fashion. Importantly, the further subjects are from each other in terms of their covariate values, the larger the variances, which reflects the underlying point that there should be greater uncertainty around estimated causal effects when there is less overlap.

The remainder of the article is organized as follows. In section 2, we formulate the Gaussian process  model and present the Bayesian inferential framework with its priors, likelihood, and posteriors. In Section 3, we conduct simulation studies to assess the performance of our method compared to other current approaches. We then apply our approach to data from an observational study of right heart catheterization in female patients in Section 4. Section 5 provides a discussion of results and offers concluding remarks.

\section{The Gaussian Process Model and Posterior Computation}\label{sec2}

\subsection{Notation and Framework for Causal Effect Estimation}

Here, we use the potential outcomes framework for estimating causal effects.\cite{rubin_2005} Suppose there are $n$ i.i.d. observations from a population. For each subject $i$ in the sample, let $A_i$ be the treatment assignment indicator with $A_i = 1$ if subject $i$ receives the treatment of interest and $A_i=0$ if subject $i$ receives the control. For a dichotomous treatment, each subject $i$ has two potential outcomes: $Y_i(1)$, the outcome under treatment, and $Y_i(0)$, the outcome under control. However, each subject may only receive one treatment in a study; that is, the observed outcome for subject $i$ is $Y_i = A_i Y_i(1)+(1-A_i) Y_i(0)$. Furthermore, define $\mathbf{X_i}$ to be a vector of $p$ pre-treatment variables or covariates. 

Our target parameter of interest is the mean difference in potential outcomes under treatment and under control, respectively, given by $\Psi=E[Y(1)-Y(0)]$. This represents the average effect had everyone been given treatment versus had everyone been given control. Here we assume that there is superpopulation of units from which the study sample is drawn, consisting of individuals who are eligible for treatment. Due to the finite size of the study sample, positivity violations can occur when certain patient characteristics are not observed in the treated sample. Identifiability of this parameter rests on the following assumptions.\cite{rosenbaum_1983, rubin_2007}
\begin{enumerate}
    \item Consistency: $Y=Y(a)$ whenever $A=a$.
    \item Ignorability: Conditional on covariates, treatment assignment is independent of the set of potential outcomes, $A \independent{\{Y(0),Y(1)\}}|X$. This essentially says that there can be no unmeasured confounding.
    \item Positivity:  The probability of receiving either treatment given the covariates is nonzero, $0<P(A=1|X)<1$.
\end{enumerate}

\subsection{Gaussian Process Model}
We assume the model for the observed continuous outcome $Y$ given confounders $X$ and treatment $A$ has the following form,\cite{hahn_2020} 
\begin{equation*}
    Y_i=\mu(X_i)+\Delta(X_i)A_i + \epsilon_i, \text{ where } \epsilon_i \sim N(0, \sigma^2).
\end{equation*}
The function $\mu(X_i)$ represents the relationship between $X$ and $Y$ that is not part of the treatment effect; that is, it is the prognostic impact of covariates. The function $\Delta(X_i)$, which is a functional coefficient of $A_i$, can be thought of as representing conditional treatment effects, reflecting interactions between covariates and treatment. Under the causal assumptions described above, the average causal effect is just $\Psi=E(\Delta(X))$.

We treat the functional form of $\mu()$ and $\Delta()$ as unknown, and therefore need to specify priors for those functions. We assume independent Gaussian process priors for these functions. Specifically, 
\begin{align*} 
    \mu(X) &\sim GP(X \beta, \mathcal{K}_{\mu}), \\
    \Delta(X) &\sim GP(0, \mathcal{K}_{\Delta}).
\end{align*}
The mean function in the prior for $\mu(X)$ centers this parameter on a linear model, $X\beta$, while the mean function for $\Delta(X)$ is zero to reflect the a priori belief of small heterogeneous treatment effects. An advantage of GP priors is that we can center the priors on a parametric model. Essentially, the prior mean based on these functions is a linear model with no effect modification. Thus, when there is limited data, the outcome model will shrink towards this prior specification.

With the goal of having a noisier mean function when there is less overlap, we choose the squared exponential (SQEXP) form for the covariance functions.
For matrices of covariate values $X=\{\mathbf{X_1}, ... , \mathbf{X_p}\}$ and $X^*=\{\mathbf{X^*_1}, ... , \mathbf{X^*_p}\}$, the covariance function $\mathcal{K}_\Delta$ is defined to be 
\begin{equation*}
    \mathcal{K}_\Delta(X,X^*|l_\Delta, \eta_\Delta) = \eta_\Delta^2\exp \left\{-\frac{1}{2}\left[ \frac{|X-X^*|}{l_\Delta}\right]^2\right\}.
\end{equation*}
The $(i,j)$th element of the covariance matrix $\mathcal{K}_\Delta$ would be
\begin{equation*}
    K_{\Delta,ij} = \mathcal{K}_\Delta(X(i), X^*(j)) = \eta_\Delta^2 \exp\left\{-\frac{1}{2}\sum_{p=1}^P \left[\frac{X_p(i)-X_p^*(j)}{l_\Delta}\right]^2\right\}.
\end{equation*}
$X_p(i)$ is the value of the $p$th covariate value for subject $i$, and $X_p^*(j)$ is the $p$th covariate value for subject $j$, $p=1,...,P$. The covariance $\mathcal{K}_\mu$ has the same form, but with different parameters, $l_\mu$ and $\eta_\mu$. The reason that this particular covariance function is useful when there are practical violations of the positivity assumption is that the variability of the function increases as distance between covariates increases, which we later show using the posterior distribution of $\Delta$. 

The hyperparameters $l$ and $\eta$ in the GP prior determine the shape and smoothness of functions defined by the prior distribution. The parameters $l_{\mu}$ and $l_{\Delta}$ are the length scales which characterize the extent to which $\mu$ and $\Delta$ function values change as the input changes.\cite{neal_1998} Small values correspond to more frequent changes in the parameter values for the same change in inputs $X$; that is, the distance in $X$ needed for the parameters to vary by an amount comparable to its range is smaller for small length scales. Larger values of this hyperparameter correspond to more smooth curves a priori. The parameters $\eta_\mu$ and $\eta_\Delta$ are the signal variances (output-scale amplitude), which control the range of the function values. For $\eta$ near 0, posterior mean estimates of the parameters $\mu$ and $\Delta$ tend to be close to each other with fewer fluctuations in the curve (closer to a straight line). At larger $\eta$ values, regions with nonoverlap will have more variability associated with the corresponding causal effect for a particular combination of covariate values. 

\subsubsection{Choice of Kernel}

The kernel or covariance function determines the types of statistical structures that may be captured by the GP model .\cite{duvenaud_2014} Our model is presented using the SQEXP kernel, resulting in functions that are infinitely differentiable to allow for smoothing. However, other stationary kernels, which only depend on the distance between two points, may also be appropriate since these methods all aim to capture the similarity in baseline characteristics of subjects.\cite{genton_2002} Common covariance function specifications include the following. \cite{rasmussen_2006}

\begin{itemize}
    \item Rational quadratic: $k(x,x') = \eta^2 \left(1 + \frac{(x-x')^2}{2\alpha l^2} \right)^{-\alpha}$.
    
    This kernel is infinitely mean square differentiable and is a scale mixture of SQEXP kernels with different length-scales. The limit of this covariance function as $\alpha \rightarrow \infty$ is the SQEXP kernel. 
    \item Mat\'ern: $k(x,x) = \eta^2 \frac{2^{1-v}}{\Gamma(v)} \left(\sqrt{2v}\frac{|x-x'|}{l} \right)^v K_v \left(\sqrt{2v} \frac{|x-x'|}{l}\right)$, where $l$ and $v$ are positive hyperparameters and $K_v$ is the modified Bessel function.\cite{abramowitz_1965, matern_1960} 
    
    This kernel also converges to the SQEXP kernel as $v \rightarrow \infty.$
    
    \item Exponential covariance function (or Ornstein-Uhlenbeck in the one-dimensional case): \\ $k(x,x') = \eta^2 exp \left (-\frac{|x-x'|}{l}\right)$.\cite{uhlenbeck_1930} 
\end{itemize}

We consider these three specifications of the covariance and compare their performances to our proposed SQEXP kernel in our simulations. In all cases, the hyperparameters are given hyperpriors such that their values are sampled in each iteration of the Markov chain Monte Carlo (MCMC) chain to allow the data to influence their values. In practice, these kernels may be combined by addition or multiplication so that the resulting kernel may capture more complexities in the data. \cite{duvenaud_2014}

\subsection{Priors, Likelihood, and Posteriors}
The outcome given treatment and confounders is distributed as $Y \sim MVN(\mu+\Delta A, \sigma^2I)$, which implies that the likelihood is  \begin{equation*}
    p(y|\mu, \Delta, \sigma^2) \propto det(\sigma^2I)^{-\frac{1}{2}} \exp \left[-\frac{1}{2}(y-(\mu+\Delta A))^T(\sigma^2I)^{-1} (y-(\mu+\Delta A))\right].
\end{equation*}
Note that $\mu=\mu(X)$ and $\Delta=\Delta(X)$ for simplification of notation. Also, $\Delta A = \begin{bmatrix} \Delta_1 A_1 \\ \vdots \\ \Delta_n A_n \end{bmatrix}$.

We specify priors for the hyperparameters
\begin{align*}
    p(\beta) &\propto MVN(0, \sigma^2_{\beta}I_P), \\
    p(l_{\mu}) & \propto gamma(l_{\mu}|\alpha_{l_{\mu}},\beta_{l_{\mu}}), \\
    p(\eta_{\mu}) & \propto gamma(\eta_{\mu}|\alpha_{\eta_{\mu}},\beta_{\eta_{\mu}}), \\
    p(l_\Delta) & \propto gamma(l_\Delta|\alpha_{l_{\Delta}},\beta_{l_{\Delta}}), \\
    p(\eta_\Delta) & \propto gamma(\eta_\Delta|\alpha_{\eta_{\Delta}},\beta_{\eta_{\Delta}}), \\
    p(\sigma^2) & \propto Inv-gamma(\sigma^2|\alpha_{\sigma^2},\beta_{\sigma^2}).
\end{align*}

The vector of coefficients $\beta$ is given a multivariate normal prior. This conjugate prior leads to a conditional posterior distribution for $\beta$ that is also multivariate normal. The hyperparameters $l$ and $\eta$ are given gamma priors since their values need to be positive. The hyperparameter $\sigma^2$ has a inverse-gamma prior, which is a common prior for variance parameters. The joint prior is $$p(\mu, \beta, l_{\mu}, \eta_{\mu}, \Delta, l_{\Delta}, \eta_{\Delta}, \sigma^2)\propto p(\mu| \beta, l_{\mu}, \eta_{\mu})p(l_{\mu}) p(\beta)p(\eta_{\mu}) p(\Delta |l_{\Delta}, \eta_{\Delta}) p(l_{\Delta})p(\eta_{\Delta})p(\sigma^2),$$ which assumes a priori independence of the hyperparameters.

Then the joint posterior is 
\begin{equation*}
    p(\mu, \beta, l_{\mu}, \eta_{\mu}, \Delta, l_{\Delta}, \eta_{\Delta}, \sigma^2|Y) \propto p(y|\mu, \Delta, \sigma^2) p(\mu| \beta, l_{\mu}, \eta_{\mu}) p(\beta) p(l_{\mu})p(\eta_{\mu}) p(\Delta |l_{\Delta}, \eta_{\Delta}) p(l_{\Delta})p(\eta_{\Delta})p(\sigma^2).
\end{equation*}

\subsubsection{Conditional Posteriors}

To ensure the Gaussian process priors for $\mu$ and $\Delta$ are not in too much disagreement with the data, we estimate hyperparameter values based on data and the posterior. That is, rather than choosing fixed values for the hyperparameters, $l$ and $\eta$, in the GP priors, we assign them gamma priors as specified in the previous subsection and use Metropolis-Hastings to update their values. These are integrated in a Metropolis within Gibbs algorithm to obtain posterior inference for $\mu$ and $\Delta$.\cite{craiu_2014}  The conditional distributions for $\beta, \mu, \text{ and } \Delta$ have analytical forms, so estimates of these parameters may be drawn directly in their corresponding Gibbs steps.\cite{gelfand_2000, casella_1992} Because these conditional distributions are needed in the algorithm and are specific to the form of our Gaussian process model, we present them here. Detailed derivations are provided in Appendix A.

\begin{itemize}
    \item $\beta|\mu, y \sim MVN \left(\left[X^T K_{\mu}^{-1}X+(\sigma^2_{\beta}I_P)^{-1}\right]^{-1}X^T K_{\mu}^{-1}\mu, \left[X^T K_{\mu}^{-1}X+(\sigma^2_{\beta}I_P)^{-1}\right]^{-1}\right)$
    \item $\mu|\beta, \Delta, y \sim MVN \left( \left[K_{\mu}^{-1}+(\sigma^2I)^{-1}\right]^{-1}\left[(\sigma^2I)^{-1}(y-\Delta A)+K_{\mu}^{-1}X\beta \right], \left[K_{\mu}^{-1}+(\sigma^2I)^{-1}\right]^{-1}\right)$
    \item To obtain the conditional posterior for $\Delta$, we first define some notation. Recall, $A$ is the vector of treatment indicators for all the subjects, and let $M$ denote a square matrix. $A^T \odot M$ indicates $A$ is multiplied element-wise to each column of $M$ while $M \odot A$ indicates $A$ is multiplied element-wise to each row of $M$.

    $\Delta|\mu, y \sim MVN\left(\left[K_{\Delta}^{-1}+A^T\odot (\sigma^2I)^{-1} \odot A \right]^{-1} A^T \odot (\sigma^2I)^{-1} (y-\mu), \left[K_{\Delta}^{-1}+A^T\odot (\sigma^2I)^{-1} \odot A\right]^{-1}\right)$
\end{itemize}

The posterior distribution for the treatment effects of all subjects has covariance matrix $[K_{\Delta}^{-1}+A^T\odot (\sigma^2I)^{-1} \odot A]^{-1}$. Because it is difficult to write out the inverses, we obtain each element for the simple case with two subjects and a single covariate X. Let $A_1=1$ and $X_1$ denote the treatment status and covariate for subject 1 and $A_2=0$ and $X_2$ denote the treatment status and covariate for subject 2, so that there is a treated subject and a control subject. The covariance matrix is given by
\footnotesize{
\begin{equation*}
    \left[K_{\Delta}^{-1} + A^T \odot (\sigma^2I)^{-1} \odot A \right]^{-1} = \begin{bmatrix} \frac{\sigma^2\eta^2_{\Delta}}{\sigma^2+\eta^2_{\Delta}}  & \frac{\sigma^2\eta_{\Delta}^2}{\sigma^2+\eta^2_{\Delta}} exp\left\{-\frac{1}{2}\left(\frac{X_1-X_2}{l_{\Delta}}\right)^2\right\} \\ \quad \quad \frac{\sigma^2\eta_{\Delta}^2}{\sigma^2+\eta^2_{\Delta}} exp\left\{-\frac{1}{2}\left(\frac{X_1-X_2}{l_{\Delta}}\right)^2\right\} & \quad \quad \eta_{\Delta}^2\left[1-\frac{\eta^2_{\Delta}}{\sigma^2+\eta^2_{\Delta}}exp\left\{-\left(\frac{X_1-X_2}{l_{\Delta}}\right)^2\right\}\right] \end{bmatrix}
\end{equation*} }

Subject 2's variance increases the further $X_2$ is from $X_1$ since a smaller amount would be subtracted from the second component of the product of $[K_{\Delta}^{-1} + A^T \odot (\sigma^2I)]^{-1}_{22}$; that is, $\left[1-\frac{\eta^2_{\Delta}}{\sigma^2+\eta^2_{\Delta}}exp\left\{-\left(\frac{X_1-X_2}{l_{\Delta}}\right)^2\right\}\right]$ becomes larger when $|X_1-X_2|$ increases. Thus, there is greater variability in the treatment effect posterior as $|X_1-X_2|$ increases. The dissimilar expressions for the diagonal elements may be attributed to $\Delta$ being the treatment effect. For a treated subject, when we condition on $\mu(X)$, we are not extrapolating when it comes to identification of $\Delta(X)$ since the mean for treated subjects is $\mu(X)+\Delta(X)$. On the other hand, for a control subject, we are extrapolating when it comes to $\Delta(X)$ because he/she was not treated--the mean for control subjects is $\mu(X)$. Because outcome information from treated subjects is driving estimation of $\Delta$, we can expect more uncertainty when estimating this parameter for control subjects. For the off-diagonal elements, larger distances in the covariates result in smaller covariances so that there is less information that can be learned from the other person, which also corresponds to larger variability.  This illustrates why with this prior specification we expect regions with little to no overlap to result in more uncertainty when it comes to causal effect estimation. 

\subsubsection{Metropolis within Gibbs Algorithm for Posterior Inference}

In this section, we briefly describe the steps of the algorithm for obtaining posterior sampling of the parameters of interest, $\mu \text{ and } \Delta$, and the hyperparameters. Details regarding the steps may be found in Appendix B. For the hyperparameters in the GP priors ($l_{\mu}, \eta_{\mu}, l_{\Delta}, \text{ and } \eta_{\Delta}$) and $\sigma^2$, we use a Metropolis-Hastings step for each one and update their values based on current values of the other parameters using an acceptance ratio. A candidate value is drawn from the proposal distribution---we employ a truncated normal distribution centered at the previous value with variance $\tau^2$, a tuning parameter, and bounded below at 0. We tune the standard deviation parameters of the proposal distribution so that the jump sizes reflect spread in the posterior and the corresponding chain trace varies quickly around the mean.\cite{gelman_2004, ellis_2018} Convergence is assessed using time-series plots for each parameter to understand the number of MCMC iterations needed to observe stabilization of chains. The acceptance ratio compares the value of the posterior at the candidate value with that at the previous value. We randomly generate a value from the standard uniform distribution $U \sim Unif(0,1)$; if $U$ is less than or equal to the ratio, than the candidate value is accepted as the parameter value at the current iteration. Otherwise, the parameter value is set to its value from the previous iteration. For $\beta, \mu, \text{ and } \Delta$, at each iteration, their new value is drawn from their respective conditional distributions given current values of all the other parameters. 

The chain is ran until the number the posterior draws after thinning and burn-ins (say, $J$) is reached. The effect estimate at iteration $j$ is obtained as the mean over the elements of the $\Delta^{(j)}$ vector (i.e., the average across all subjects): $\psi^{(j)} = \frac{1}{n} \sum_{i=1}^n \Delta^{(j)}_i$. The average treatment effect is then estimated as the mean of the posterior draws of $\psi$, $\Psi = \frac{1}{J}\sum_{j=1}^J \psi^{(j)}$, so that estimation for continuous outcomes may be obtained directly from the posterior distribution of $\Delta$.

\subsection{Extension to Binary Outcomes}
In this section, we extend our model to dichotomous outcomes where $Y$ may take on the values of either 0 or 1. Let $\theta=\{\mu, \beta, l_{\mu}, \eta_{\mu}, \Delta, l_{\Delta}, \eta_{\mu}\}$. The probit model assigns to each $X_i$ the variable $Y_i \in \{0,1\}$ using $P(Y_i=1)=\Phi(f(X_i,A_i, \theta))=\Phi(\mu(X_i)+\Delta(X_i) A_i)$, where $\Phi$  is the standard normal cumulative distribution function. Assuming the same priors for the parameters in $\theta$ as those for the continuous outcome case, the posterior is 
\begin{equation*}
    p(\theta|X,A,Y) \propto p(Y|X, A,\theta) p(\theta) \propto \prod_{i=1}^n \Phi(f(X_i,A_i, \theta))^{Y_i}(1-\Phi(f(X_i,A_i,\theta))^{1-Y_i} p(\theta)
\end{equation*}

Sampling $\theta$ from this form is difficult. Thus, we consider the model augmented with a random variable $Z$.\cite{meng_1999, van_dyk_2001} Specifically, we define independent latent variables $Z_i$, where each $Z_i$ is normally distributed. Then the augmented probit model has the hierarchical structure as follows:
\begin{align*}
    Y_i &=\begin{cases}
    1, & \text{if } Z_i > 0 \\
    0, & \text{if } Z_i \leq 0
    \end{cases} \\
    Z_i|\theta, X_i &= \mu_i+\Delta_i A_i + \epsilon_i, \epsilon_i \sim N(0,1) \\
    \theta &\sim p(\mu, \beta, l_{\mu}, \eta_{\mu}, \Delta, l_{\Delta}, \eta_{\mu})
\end{align*}

Here, $Y_i$ is deterministic conditional on the sign of $Z_i$.\cite{albert_1993} Under the augmented model, $P(Y_i=1|X_i,A_i,\theta)= \Phi(f(X_i,A_i, \theta))$,
so the two models give the same inference. We will employ the augmented model in sampling of the parameters of interest and the latent variables. The joint posterior of latent variables $Z$ and model parameters $\theta$ given data $X,A,Y$ is
\begin{align*}
    p(Z,\theta|X,A,Y) & \propto p(Z|X, A, Y, \theta) p(\theta)  \\
    &\propto p(Z|\mu, \Delta, A, X, Y)p(\mu|\beta, l_{\mu}, \eta_{\mu}, X)p(\beta)p(l_{\mu})p(\eta_{\mu}) p(\Delta|l_{\Delta}, \eta_{\Delta}) p(l_{\Delta})p(\eta_{\Delta})
\end{align*}
where
\begin{align*}
    p(\mu_i|\beta, l_{\mu}, \eta_{\mu}, X) &= N(\mu|X\beta, \mathcal{K}_{\mu}) \\
    p(\Delta|l_{\Delta}, \eta_{\Delta}, X) &= N(\Delta|0, \mathcal{K}_{\Delta}) \\
    p(Z|\mu, \Delta, A, Y) &= N(Z|\mu+\Delta A, 1) [I(Y=1)I(Z>0)+I(Y=0)I(Z\leq 0)]
\end{align*}

Note that $\theta$ is not dependent on $Y$ given $Z$, so the conditional posterior of the model parameters $\theta$ is
\begin{equation*}
    p(\theta|Z,X) \propto p(\theta) N(Z|\mu+\Delta A,1).
\end{equation*}

The conditional posterior of the latent variable $Z_i$ is
\begin{equation*}
    Z_i|\theta, Y_i, X_i \sim \begin{cases}
    TN(mean=\mu_i+\Delta_i A_i, sd=1, lower=0, upper=\infty), & \text{if } Y_i=1 \\
    TN(mean=\mu_i+\Delta_i A_i, sd=1, lower=-\infty, upper=0), & \text{if } Y_i=0
    \end{cases}
\end{equation*}

Estimates of parameters are obtained by modifying the Metropolis within Gibbs algorithm such that $Z_i$ takes the place of the continuous outcome and an additional step is used to sample $Z_i$ from a truncated normal distribution.

Our interest is in  the causal risk difference,
$\Psi=P\{Y(1)=1\}-P\{Y(0)=1\}$. 
The posterior for $\Psi$ can be obtained as follows. From the Gibbs sampler, we will have stored $J$ draws of $\mu$ and $\Delta$
(after discarding burn-ins and thinning). At each iteration $j$, we obtain a draw of each potential outcome via computation. The probability of outcome under treatment, $p_1^{(j)}=P(Y(1)=1)^{(j)}$, is
\begin{equation*}
    p_1^{(j)} = \frac{1}{n} \sum_{i=1}^n \Phi(\mu_i^{(j)}+\Delta_i^{(j)})
\end{equation*}
and the probability of the outcome in the absence of treatment is, $p_0^{(j)}=P(Y(0)=1)^{(j)}$, is
\begin{equation*}
    p_0^{(j)} = \frac{1}{n} \sum_{i=1}^n \Phi(\mu_i^{(j)}).
\end{equation*}
Then the effect estimate at iteration $j$ of the MCMC chain is $\Psi^{(j)} = p_1^{(j)}-p_0^{(j)}$.
The estimate of the risk difference is the average difference in proportions over the $J$ posterior samples: $\Psi=\frac{1}{J} \sum_j^{J} \Psi^{(j)}$.

\section{Simulation Studies}\label{sec3}

Simulation studies were used to assess the performance of the GP model for scenarios with varying degrees of nonoverlap. In these, we considered both linear and nonlinear response surfaces with the latter including treatment heterogeneity and interactions between covariates. We compared our GP approach to the following methods:

\begin{description}
\item[BCF] Bayesian causal forest with the prognostic and treatment components as functions of covariates and propensity scores, as proposed by Hahn et al. (2020).\cite{hahn_2020}
\item[BART-Stratified] separate BART models are fit to treated and control subjects using covariates only and potential outcomes are estimated as the expected value of the function.\cite{chipman_2010}
\item[BART-Single] untrimmed BART as implemented by Nethery et al. (2019) in which the treatment variable and propensity score are included as covariates, and potential outcomes are estimated with posterior predictive distributions.\cite{chipman_2010, nethery_2019} A single model is fit for the entire sample.
\item[BART+SPL] the method proposed by Nethery et al. (2019) for nonoverlap using the recommended $a=.1, b=10$ to define the region of overlap based on propensity scores.\cite{nethery_2019}
\item[GLM] generalized linear model regression of outcome on main effects of treatment indicator and covariates with identity link for continuous outcomes and probit link for binary outcomes.
\end{description}

For the Bayesian methods, MCMC specifications include 10,000 burn-ins and 5000 iterations after burn-ins, in which every 5th is kept, yielding 1000 posterior draws of the average treatment effect. 

We use 1000 replications for the simulations. For each simulated data set, we obtain 1000 posterior estimates (after discarding burn ins and thinning) of treatment effect by averaging over all subjects. Specifically, for each replication $k$, we have 1000 posterior draws of the treatment component (average over individual causal effects at each iteration). Denote the estimate of the treatment effect, the mean over the posterior draws, by $\Psi_k$. The standard deviation $SD_k$ and 95\% credible intervals $CI_k$ are obtained from these 1000 posterior draws. Over the 1000 replications we compute several quantities to measure performance: 
\begin{align*}
    ATE &= \frac{1}{1000} \sum_{k=1}^{1000} \Psi_k
    & Bias &= \frac{1}{1000} \sum_{k=1}^{1000} (\Psi_k - ATE_{true}) \\
    \text{\% Bias} &= \frac{1}{1000} \sum_{k=1}^{1000} \frac{\Psi_k-ATE_{true}}{|ATE_{true}|}\cdot 100 &
    \overline{SD} &= \frac{1}{1000} \sum_{k=1}^{1000} SD_k \\
    SE &= \sqrt{\frac{1}{1000-1}\sum_{k=1}^{1000} (\Psi_k-ATE)^2}  & MSE &= \frac{1}{1000} \sum_{k=1}^{1000} (\Psi_k - ATE_{true})^2 \\
    Coverage &= \frac{1}{1000} \sum_{k=1}^{1000} I(ATE_{true} \in CI_k)
\end{align*}

These simulations were conducted in R (R Core Team, 2021).\cite{r_2021}

\subsection{Continuous Outcome}

For the setting with a continuous outcome, we generate treatment indicator $A$, two continuous covariates, and one binary covariate for $n=500$ subjects. Our data generating model is specified as follows. First, treatment status is simulated as $A \sim Bernoulli(.5)$, and covariates are simulated based on treatment received. If $A=1$, the covariates are generated as $X_1 \sim N(\mu_1,1), X_2 \sim N(\mu_2,1), X_3 \sim Bernoulli  (p)$; if $A=0$, $X_1 \sim N(0,1), X_2 \sim N(2,1), X_3 \sim Bernoulli (.4)$.
We consider two different outcome models. 
\begin{enumerate}
    \item $Y_1 \sim N(1-2X_1+X_2-1.2X_3+2A,1)$
    \item $Y_2 \sim N(-3-2.5X_1+2X_1^2 A+exp(1.4-X_2 A)+X_2X_3-1.2X_3-2X_3 A + 2A, 1)$
\end{enumerate}

Different combinations of $\mu_1$, $\mu_2$, and $p$ are chosen to control the amount of covariate nonoverlap in the sample. We consider settings with some nonoverlap $(\mu_1=1, \mu_2=2, p=.5)$ and substantial nonoverlap $(\mu_1=1, \mu_2=3, p=.6)$. In the first model, the outcome is linearly related to covariates and treatment. In this case, for any combination of covariates, the treatment effect is the same--that is, treatment effect is homogeneous. We expect all methods to have decent performance since there are no interactions between covariates and treatment in the outcome model. Data generating model 2 incorporates nonlinearity and treatment heterogeneity. For instance, as $X_1$ values increase, the outcome $Y$ for treated subjects tends to increase while $Y$ values for control subjects tends to decrease. This leads to treatment effects that are larger in magnitude at larger $X_1$ values. Further, there is nonoverlap for these $X_1$ values since only treated subjects are observed in this region. The combination of treatment heterogeneity and nonoverlap makes it difficult to assess the treatment effect; parametric models would find it especially challenging to capture the true relationships. 

We also consider the set of simulation scenarios in  Nethery et al. (2019) since our primary comparative method is BART+SPL.\cite{nethery_2019} We use scenarios for which  BART+SPL had the best performance; this method underperforms when propensity scores are misspecified as shown by simulation results from Nethery et al. (2019) that considered misspecified propensity score models.\cite{nethery_2019}. Specifically, they show that larger bias and lower coverage resulted when estimated propensity scores from logistic regression are used. For $n=500$ subjects, half are assigned treatment $A=1$ and half are assigned to $A=0$. Here, $c$ controls the degree of overlap. The values of $c$ that are considered are $0, 0.35, 0.7$, where larger values correspond to greater extents of nonoverlap. Covariates are generated based on treatment assignment. For treated subjects ($A=1$), the covariates are generated with $X_1 \sim Bernoulli(.5), X_2 \sim N(2+c, (1.25 + .1c)^2)$. For control subjects ($A=0$), $X_1 \sim Bernoulli(.4), X_2 \sim N(1,1)$. The true propensity score is calculated based on density functions as follows: 
\begin{equation*}
    \text{True PS} =\frac{N(X_2; \mu=2+c, \sigma=1.25+.1c)\cdot Ber(X_1;p=.5)}{N(X_2; \mu=2+c, \sigma=1.25+.1c)\cdot Ber(X_1;p=.5)+N(X_2; \mu=1, \sigma=1)\cdot Ber(X_1;p=.4)}
\end{equation*}

The true potential outcomes under control and under treatment for all subjects are generated: 
\begin{align*}
    Y(0) &= -1.5X_2,\\
    Y(1) &= \frac{-3}{1+\exp(-10(X_2-1))}+.25X_1-X_1X_2.
\end{align*}
Then the true treatment effect for each person is $Y_i(1)-Y_i(0)$, so that there is a ``true'' ATE value for each simulated data set (say $ATE_{true,k}$ for the $k$th replication).
\begin{equation*}
    ATE_{true,k} = \frac{1}{N}\sum_{i=1}^N Y_i(1)-Y_i(0)
\end{equation*}
The observed outcome is taken to be $Y_i=A_i Y_i(1) + (1-A_i) Y_i(0)$. 

Because there is a true ATE value for each replication indexed by $k$, for $k=1,...,1000$, we modify the bias and coverage metrics to accommodate different true values across the replications.
\begin{align*}
    \text{Bias} &= \frac{1}{1000} \sum_{k=1}^{1000} (\Psi_k-ATE_{true,k}) \\
    \text{\% Bias} &= \frac{1}{1000} \sum_{k=1}^{1000} \frac{\Psi_k-ATE_{true,k}}{|ATE_{true,k}|} \cdot 100 \\
    \text{Coverage} &= \frac{1}{1000} \sum_{k=1}^{1000} I(ATE_{true,k} \in CI_k).
\end{align*}

\subsubsection{Results}
Simulation results for the nonoverlap scenarios in which the continuous outcomes are generated with a linear response surface are presented in Table 1. All methods provide estimates with low bias except BART-Stratified in the setting with some nonoverlap. We observe differences in the variability of the estimates. In particular, in the setting with some nonoverlap, the GP model's estimate of variability is closest to that obtained from linear regression (the gold standard in this case). As the extent of nonoverlap increases, the variability obtained from the GP model increases to account for the greater uncertainty in those regions while maintaining nominal coverage. On the other hand, with increasing amounts of nonoverlap, the BART+SPL method results in doubled MSE and coverage very close to 1 (indicating overcoverage). 

\begin{center}
\begin{table}[H]
\caption{Effect estimates for nonoverlap scenarios involving a linear response surface across methods. The true ATE is 2 for both degrees of nonoverlap.\label{tab:Results2L}}
\centering
\begin{tabular}{c|c|ccccccc}
\hline
                                                                                  & Method          & ATE   & Bias  & \% Bias & $\overline{SD}$ & SE   & MSE  & Coverage \\ \hline
\multirow{6}{*}{\begin{tabular}[c]{@{}c@{}}Some\\ nonoverlap\end{tabular}}        & GP              & 1.968 & -.032 & -1.604  & .102                             & .100 & .011 & .948     \\
                                                                                  & BCF             & 2.002 & .002  & .082    & .112                             & .104 & .011 & .961     \\
                                                                                  & BART-Stratified & 1.904 & -.096 & -4.820  & .111                             & .113 & .022 & .855     \\
                                                                                  & BART-Single     & 2.011 & .011  & .561    & .115                             & .106 & .011 & .968     \\
                                                                                  & BART+SPL        & 2.015 & .015  & .729    & .156                             & .118 & .014 & .980     \\
                                                                                  & Linear model    & 1.998 & -.002 & -.110   & .101                             & .098 & .010 & .945     \\ \hline
\multirow{6}{*}{\begin{tabular}[c]{@{}c@{}}Substantial\\ nonoverlap\end{tabular}} & GP              & 1.971 & -.029 & -1.458  & .115                             & .110 & .013 & .947     \\
                                                                                  & BCF             & 1.971 & -.029 & -1.428  & .130                             & .118 & .015 & .962     \\
                                                                                  & BART-Stratified & 1.946 & -.054 & -2.684  & .136                             & .133 & .020 & .927     \\
                                                                                  & BART-Single     & 1.985 & -.015 & -.772   & .137                             & .122 & .015 & .965     \\
                                                                                  & BART+SPL        & 1.976 & -.024 & -1.203  & .288                             & .175 & .031 & .997     \\
                                                                                  & Linear model    & 1.999 & -.001 & -.070   & .112                             & .108 & .012 & .955     \\ \hline
\end{tabular}
\end{table}
\end{center}

Simulation results for data simulated using the second data generating model are given in Appendix C. For both degrees of nonoverlap, the GP model provided estimates with the smallest bias among the comparator methods. For this complex data scenario, linear regression has the worst performance as expected. The other nonparametric models, BCF, the two BART models, and BART+SPL, all underestimate the average treatment effect with BART+SPL being the most biased. Note that the variability in the estimates of average treatment effect from the GP model again increase as the amount of nonoverlap increases as shown by $\overline{SD}$ values. Thus, although there was bias from all the methods, the GP performed the best in terms of bias and efficiency, as reflected in its MSE metric, among the approaches considered. 

Table \ref{tab:NetheryResults} presents the simulation results using the scenarios from Nethery et al. (2019). For each degree of nonoverlap, the GP model results in the smallest bias---even under the settings where BART+SPL was previously shown to perform best.\cite{nethery_2019} The GP model has coverage that is nearly the same as that of  BART+SPL but has variability estimates that are smaller. The high coverage for the GP model indicates that it both accurately estimates the truth and translates the uncertainty from nonoverlap regions into higher variability. The greater than nominal coverage from the GP model for these scenarios may be due to the absence of an error term when outcomes were generated. In this scenario, estimates provided by the GP model had smaller bias than estimates from BCF. The increase in the variability estimates as the amount of nonoverlap grows is larger for the GP model than for the BART-only models (BCF, BART-Stratified, and BART-Single), better reflecting the extent of nonoverlap. Given the nonlinearity and interactions specified in the outcome model, the parametric linear regression has the worst performance as expected. 

\begin{table}[H]
\caption{Performance of the methods for nonoverlap scenarios from Nethery et al. (2019) that employ true propensity scores.}
\label{tab:NetheryResults}
\centering
\begin{tabular}{c|c|ccccccc}
\hline
Setting                & Method          & ATE   & Bias  & \% Bias   & $\overline{SD}$ & SE   & MSE                    & Coverage \\ \hline
\multirow{6}{*}{c=0}   & GP              & -.264 & .001  & .326      & .024                             & .051 & $8.916 \times 10^{-5}$ & 1.000    \\
                       & BCF             & -.296 & -.031 & -12.491   & .012                             & .057 & .002                   & .369     \\
                       & BART-Stratified & -.304 & -.039 & -15.643   & .014                             & .056 & .002                   & .378     \\
                       & BART-Single     & -.287 & -.022 & -8.981    & .050                             & .056 & .001                   & .990     \\
                       & BART+SPL        & -.256 & .009  & 3.486     & .063                             & .054 & $3.216 \times 10^{-4}$ & 1.000    \\
                       & Linear model    & -.330 & -.065 & -26.642   & .082                             & .073 & .008                   & .941     \\ \hline
\multirow{6}{*}{c=.35} & GP              & -.190 & -.002 & -1.583    & .030                             & .057 & $2.910 \times 10^{-4}$ & .998     \\
                       & BCF             & -.242 & -.054 & -33.446   & .016                             & .067 & .004                   & .257     \\
                       & BART-Stratified & -.263 & -.075 & -45.849   & .020                             & .067 & .007                   & .204     \\
                       & BART-Single     & -.231 & -.043 & -26.700   & .054                             & .068 & .004                   & .905     \\
                       & BART+SPL        & -.173 & .015  & 8.714     & .073                             & .060 & $7.744 \times 10^{-4}$ & 1.000    \\
                       & Linear model    & -.363 & -.175 & -110.756  & .093                             & .086 & .037                   & .546     \\ \hline
\multirow{6}{*}{c=.70} & GP              & -.095 & -.009 & -54.992   & .041                             & .066 & $9.488 \times 10^{-4}$ & .995     \\
                       & BCF             & -.172 & -.087 & -333.973  & .020                             & .081 & .011                   & .189     \\
                       & BART-Stratified & -.215 & -.130 & -472.294  & .026                             & .083 & .021                   & .123     \\
                       & BART-Single     & -.158 & -.073 & -282.576  & .058                             & .085 & .010                   & .705     \\
                       & BART+SPL        & -.058 & .026  & 72.858    & .087                             & .070 & .002                   & .998     \\
                       & Linear model    & -.402 & -.316 & -1128.419 & .106                             & .102 & .110                   & .137     \\ \hline
\end{tabular}
\end{table}

\subsubsection{Illustration of Individual Causal Effect Estimates}
\begin{figure}[H]
    \centering
    \includegraphics[width=14cm]{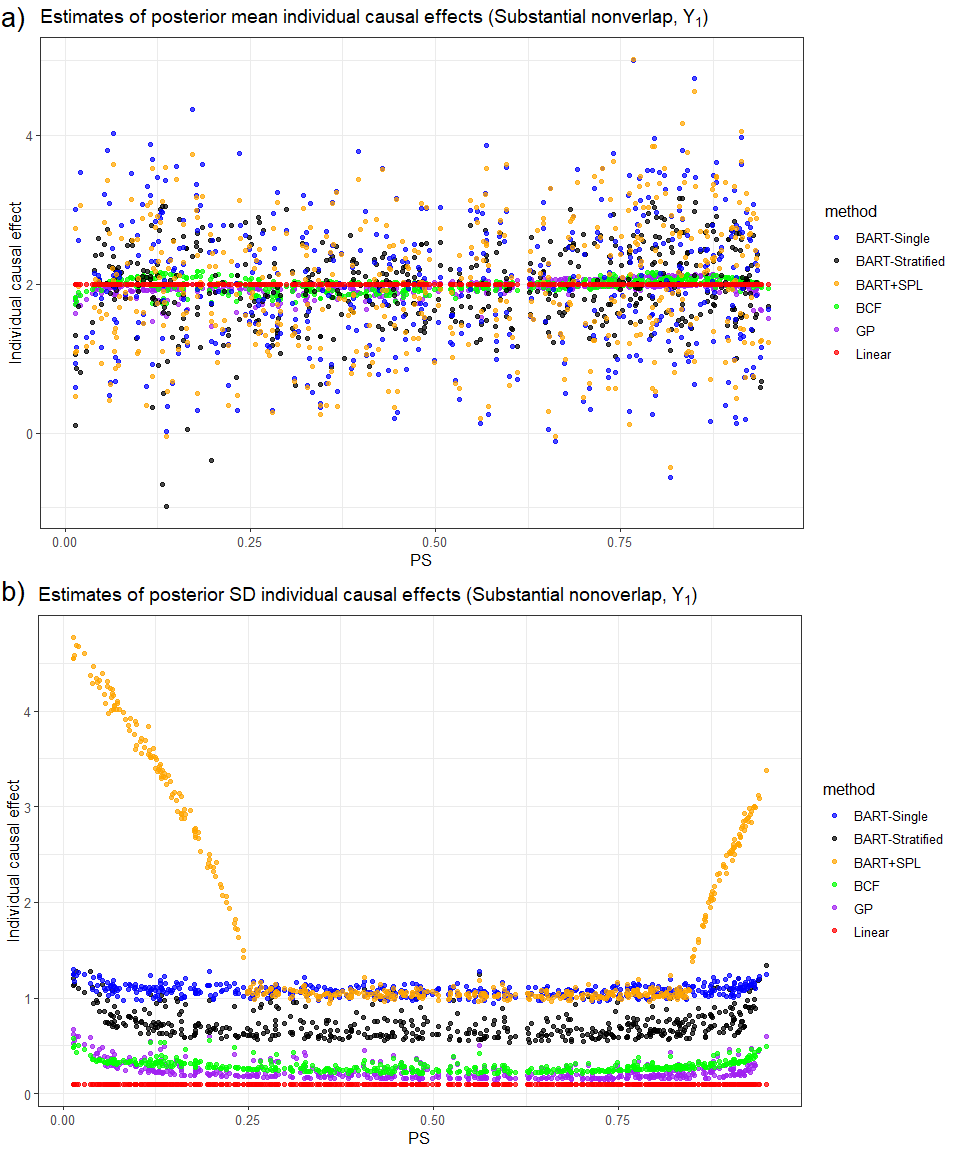}
    \caption{Individual causal effect exploration when outcome is generated with $Y_1$ for the substantial nonoverlap case.}
    \label{fig:L-ICE}
\end{figure}

To illustrate \textit{individual} causal effects obtained by each method using, we use one simulated data for each nonoverlap scenario. For the linear response surface under substantial nonoverlap, we plot estimates of posterior mean and posterior standard deviation of individual causal effects across the iterations for each subject (Figure \ref{fig:L-ICE}). The analogous figure for the moderate nonoverlap setting is included as Supplementary Figure S1. The GP model provides estimates of individual causal effects that are close to 2 (the truth) indicating high precision, while those provided by BART-Single and BART+SPL are highly variable. Although the average treatment effect given by BART+SPL is close to the true value, the estimates of individual treatment effects range from -0.36 to 4.98, reflecting great variability.  

The GP model tends to provide the lowest posterior standard deviations in regions of overlap. These estimates are only slightly greater than estimates of variability from linear regression, which would be correctly specified for this scenario. Because BART-Stratified models the treatment and control groups separately, baseline variability in estimates of treatment effect is higher. Notably,  the Bayesian approaches show increases in variability in areas of increasing non-overlap, indicating that they are capturing the greater uncertainty in those regions. The GP model's estimates of individual-level standard deviations are greater than those of BCF in the nonoverlap areas; the continuous nature of the GP model means that uncertainty increases with covariate distance. However, the increase in the posterior SD values when using  BART+SPL is so steep that estimates for individuals in the nonoverlap areas may hold little value.

\begin{figure}[H]
    \centering
    \includegraphics[width=13.5cm]{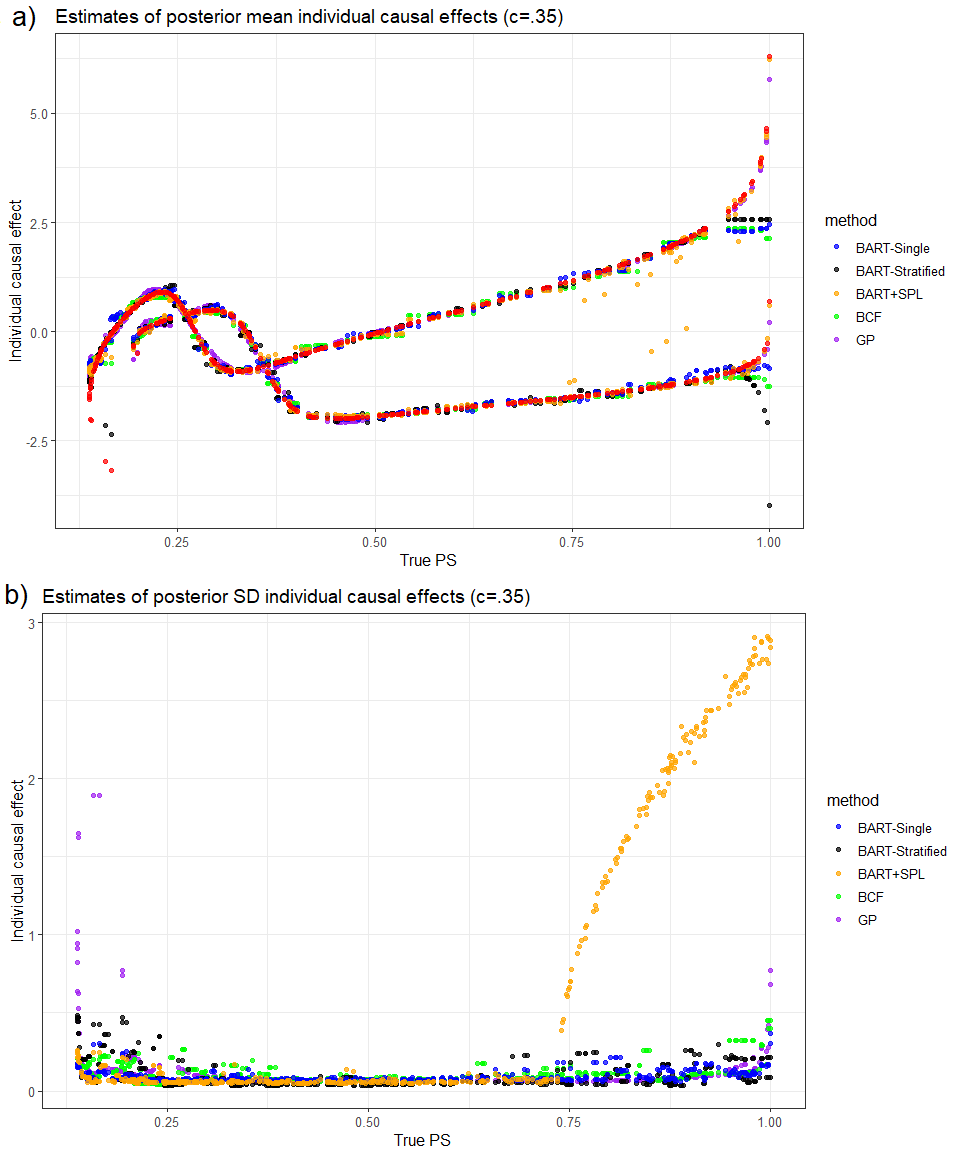}
    \caption{Individual-level posterior mean and standard deviation estimates from the methods considered. Red points denote the true individual causal effect based on the data generating model.}
    \label{fig:c.35ICE}
\end{figure}

In Figure \ref{fig:c.35ICE}, we display individual causal effects when there is moderate overlap (c=.35). The plots for c=0 and c=.70 are similar and are included as Supplementary Figures S2 and S3, respectively. In this scenario, it is clear that the BART-only methods tend to give constant estimates in regions of nonoverlap (due to splitting on tails of covariate distributions) while the GP model and the BART+SPL model are able to capture the trends since their estimates of individual-level effects are closer to the true values. Thus, the BART-only methods underestimate the treatment effect in these areas. BART+SPL deems those with propensity scores larger than .75 to be in the region of nonoverlap, so that the posterior standard deviations increase substantially for these subjects. The increase in individual level posterior standard deviations for the GP model is more gradual. Further, the few treatment subjects relative to the number of control subjects with PS near 0 is reflected in the larger variability as estimated by the GP model for these regions. BART+SPL does not take this into account for the specified $a$ and $b$ values.

\subsubsection{Sensitivity to Specification of the Gaussian Process Prior}
In the proposed GP model, prior distributions are specified for the hyperparameters instead of setting them to specific values in an effort to reduce sensitivity to these specifications. Further, a choice must be made about the covariance function $\mathcal{K}$. We explore the sensitivity of the model to these choices using two representative simulation scenarios--the some nonoverlap scenario with outcome generating model $Y_1$ and the nonoverlap setting from Nethery et al. (2019) with $c=.35$. The first one involves a small degree of nonoverlap and a linear response surface while the second one involves a moderate degree of nonoverlap and a nonlinear response surface, reflecting varying data complexity. 

For our primary implementation of the Gaussian process prior, we employed gamma (2,1) as the hyperprior for the $l$ and $\eta$ parameters--that is, we set $\alpha_{l_\mu}, \alpha_{\eta_{\mu}}, \alpha_{l_{\Delta}}, \alpha_{\eta_{\Delta}} = 2$ and $\beta_{l_\mu}, \beta_{\eta_{\mu}}, \beta_{l_{\Delta}}, \beta_{\eta_{\Delta}} = 1$. These values corresponds to a distribution that is right skewed and gives large probabilities to values near 0. To see if the model is sensitive to these hyperprior specifications, we examine the performance when $l$ and $\eta$ are assumed to have a $gamma  (4, .5)$ prior, such that $\alpha_{l_\mu}, \alpha_{\eta_{\mu}}, \alpha_{l_{\Delta}}, \alpha_{\eta_{\Delta}} = 4$ and $\beta_{l_\mu}, \beta_{\eta_{\mu}}, \beta_{l_{\Delta}}, \beta_{\eta_{\Delta}} = .5$. This distribution is centered at larger values and has a much larger spread (Supplementary Figure 4). Simulation results comparing these two specifications are given in Table 1 in Appendix D. Measures of bias, variability, and coverage are similar for these different hyperprior specifications. 

Next, we compare different specifications of the covariance function for these two representative simulation settings. We compare the SQEXP covariance function to the rational quadratic, Mat\'ern, and Ornstein-Uhlenbeck (exponential) covariance functions. For these specifications, the hyperparameters are given a gamma (2,1) prior and sampled using MCMC. Results are provided in Table 2 in Appendix D. Again, estimates are in the same ballpark across the different covariance functions specifications, suggesting that the model is relatively insensitive to which kernel is used.  

\subsection{Binary Outcomes}
In simulation studies for binary outcomes, the treatment indicator $A$ and covariates $X_1, X_2,$ and $X_3$ are generated identically to the continuous case. We again consider two levels of nonoverlap and two outcome generating distributions. Outcomes are drawn from the Bernoulli distribution with proportion parameter that depends on covariate values and treatment received.
\begin{itemize}
    \item $Y_{1B} \sim Bernoulli(\Phi(-1-2X_1+X_2-1.2X_3+2A))$
    \item $Y_{2B} \sim Bernoulli(\Phi(-3-2.5X_1+2X_1^2 A+exp(1.4-X_2A)+1X_2X_3-1.2X_3-2X_3 A + A))$
\end{itemize}

Detailed results of these simulations are provided in Appendix E. In brief, the GP model gives lower or similar bias and higher efficiency compared to the other methods and maintains its coverage when nonlinear and interaction terms are added to the outcome models. 

\subsection{High Dimensional Covariate Setting}
Covariate nonoverlap is more likely to occur when there are many variables that are controlled for. To explore this setting, we simulate data according to the high dimensional scenarios employed in Nethery et al. (2019) and compare the performance of the GP model to our primary comparator, the BART+SPL method.\cite{nethery_2019} Implementation of the BART+SPL approach follows the original procedure used by Nethery et al. when considering the high dimensional setting. Again, half of the $N=500$ subjects are assigned to treatment (A=1) and the other half are placed in the control group (A=0). 10 confounders (variables associated with both treatment and outcome) are generated. For treated subjects ($A=1$), the covariates are generated with $X_1,..., X_5 \sim Bernoulli(.45), X_6,...,X_{10} \sim N(2, 4)$. For control subjects ($A=0$), $X_1,...,X_5 \sim Bernoulli(.4), X_6,...,X_{10} \sim N(1.3,1)$. The true potential outcomes under control and under treatment for all subjects are generated as follows. 
\begin{align*}
    Y(0) &= .5(X_1+X_2+X_3+X_4+X_5)+15(1+exp(-8X_6+1))^{-1}+X_7+X_8+X_9+X_{10}-5,\\
    Y(1) &= X_1+X_2+X_3+X_4+X_5-.5(X_6+X_7+X_8+X_9+X_{10})
\end{align*}

Three scenarios are explored via simulations.
\begin{enumerate}
    \item HD 1: Only the 10 confounders are included in the model.
    \item HD 2: The 10 confounders and 25 additional randomly generated variables (not truly related to the outcome variable) are included in the model.
    \item HD 3: The 10 confounders and 50 additional variables are included in the model. 
\end{enumerate}

\begin{table}[H]
\centering
\caption{Comparisons of performance of the GP model to the BART+SPL method for high-dimensional covariate settings.}
\label{tab:HD}
\centering
\begin{tabular}{cc|ccccccc}
\hline
                      &          & ATE     & Bias  & \% Bias & $\overline{SD}$ & SE    & MSE   & Coverage \\ \hline
\multirow{2}{*}{HD 1} & GP       & -16.688 & .732  & 4.203   & .353            & .513  & .684  & .430     \\
                      & BART+SPL & -17.631 & -.212 & -1.218  & .667           & 1.877 & 3.462 & .524     \\ \hline
\multirow{2}{*}{HD 2} & GP       & -18.315 & -.912 & -5.249  & .272            & .380  & .896  & .072     \\
                      & BART+SPL & -16.991 & .412  & 2.371   & .537           & 1.712 & 2.961 & .492     \\ \hline
\multirow{2}{*}{HD 3} & GP       & -18.125 & -.715 & -4.113  & .300            & .405  & .588  & .354     \\
                      & BART+SPL &  -16.856  & .554  & 3.169        &   .513   &  1.548  & 2.671 & .486  \\ \hline
\end{tabular}
\end{table}

Measure of bias, variability, and coverage for the GP model and the BART+SPL method are presented in Table \ref{tab:HD}. Bias is slightly higher from the GP model for these particular simulation scenarios. However, the MSE from the GP model is less than that from the BART+SPL method. While the bias and coverage from the BART+SPL method tends to get worse as additional variables are included in the model, there is no clear pattern for the GP model. Further investigation of the GP model's performance and modification to its form or prior specifications may be needed to better accommodate high dimensional covariate settings.

\section{Application to Study of Right Heart Catheterization} \label{sec4}
We applied our GP approach to data on critically ill patients in the Study to Understand Prognoses and Preferences for Outcomes and Risks of Treatments (SUPPORT). Details on the study population and data collection have been previously described in Connors et al. (1996).\cite{connors_1996} We note that the purpose of this data example is to demonstrate our method and not to make clinical claims. These data are publicly available which will allow readers to readily replicate our results. In our analysis, we assess the effects of right heart catheterization (RHC) in the first 24 hours upon entry into study on survival for female subjects. The binary outcome of interest in this study is defined as \begin{equation*}
    Y_i=
    \begin{cases}
    1, & \text{ if subject } i \text{ died within 180 days} \\
    0, & \text{ otherwise}
    \end{cases}
\end{equation*}

\begin{figure}[H]
    \centering
    \includegraphics[width=13cm]{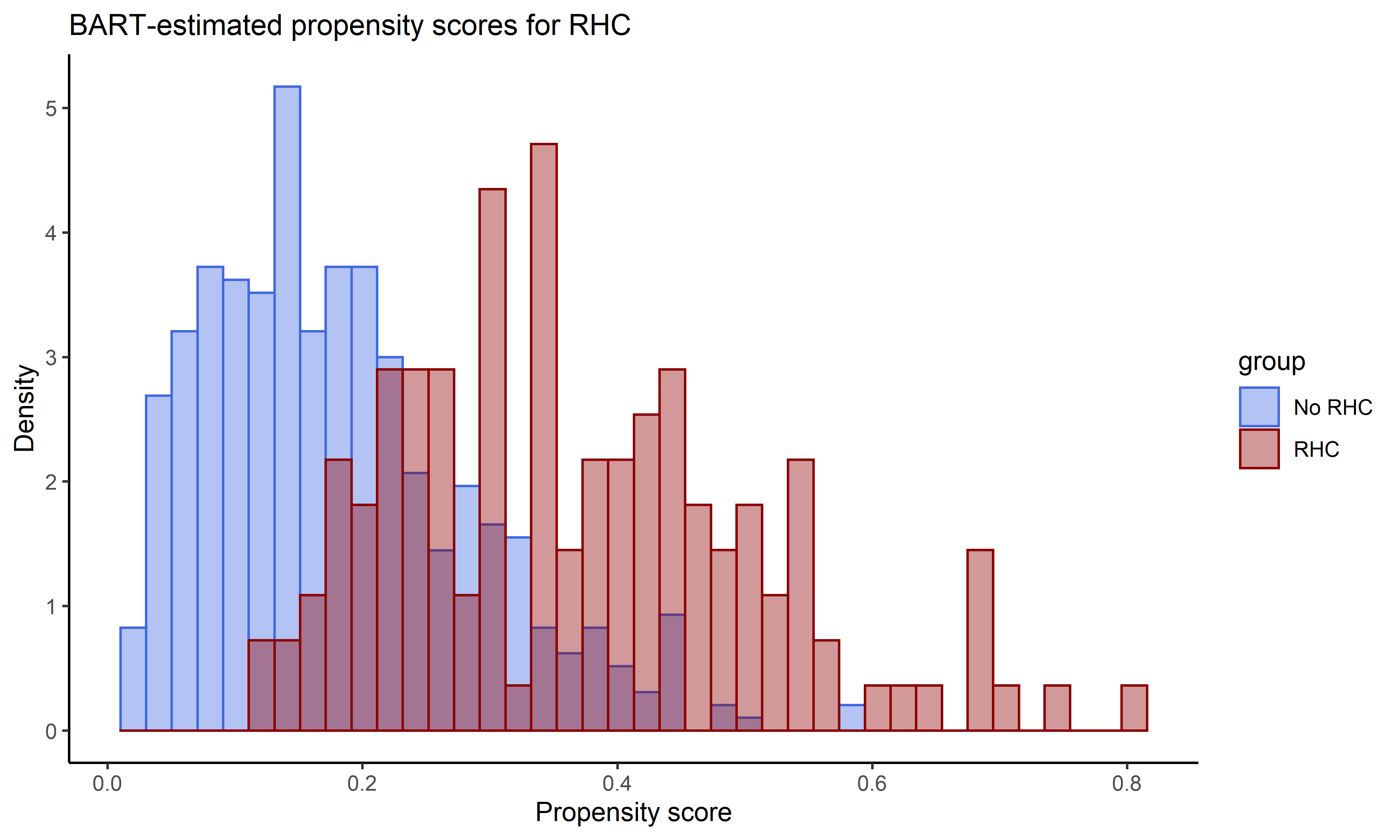}
    \caption{Histograms of estimated propensity scores for patients who received an RHC and those that did not.}
    \label{fig:RHC_PS}
\end{figure}

The confounding variables of interest include age, race, years of education, income, medical insurance, primary disease category, Activities of Daily Living score, Duke Activity Status Index, do-not-resuscitate status, cancer status, SUPPORT model estimate of the probability of surviving 2 months, APACHE III score, coma score based on Glasgow on day 1, physiological measurements, and categories of comorbid illness. Of the 617 female subjects of interest, 137 received an RHC while 480 did not. In the treatment group, 62 died within 180 days, compared with 219 in the control group. Characteristics of the sample for analysis are provided in Appendix F. We see that the RHC group tends to be younger on average, have higher income, are less likely to sign a do-not-resuscitate form, and have lower respiratory rates and lower PaCO2 on Day 1. Further, the proportions of people with pulmonary disease were significantly different between the RHC and non-RHC groups.  Propensity scores were estimated using a BART probit model with treatment status as the response and all confounding variables as predictors. Figure \ref{fig:RHC_PS} demonstrates nonoverlap in the tails. 
 
To fit the Gaussian process model, we employed four chains with different initial values to estimate the parameters of interest. Specifically, each chain involved 10,000 burn-ins and 20,000 iterations after burn-ins, from which every 80th was kept in order to minimize autocorrelation. Combining these iterations, we obtained 1000 posterior draws for the parameters. We calculated a risk difference defined as the mean difference in probabilities of dying within 180 days from the start of study entry had the RHC been given versus had the RHC not been given, respectively. The risk difference estimate was 0.024 with   95\% credible interval [-0.031, 0.098]. These estimates indicate that the 180-day survival of subjects who received the RHC did not differ significantly from that of patients who did not get RHC. Point estimates obtained from the comparator methods were found to be similar as shown in Table \ref{tab:dataResults}. The GP model resulted in  narrower credible intervals, which is consistent with what we found in some of the simulation studies (Tables \ref{tab:Results2L} and \ref{tab:NetheryResults}). 

\begin{table}[h]
\caption{Estimated average treatment effect of receiving the RHC.}
\label{tab:dataResults}
\centering
\begin{tabular}{cccc}
\hline
                & ATE   & SE   & 95\% CrI        \\ \hline
GP              & .022 & .032 & [-.032, .100] \\
BCF             & .033 & .044 & [-.047, .126] \\
BART-Stratified & .035 & .049 & [-.059, .124] \\
BART-Single    & .031 & .042 & [-.049, .114] \\
BART+SPL        & .030 & .077 & [-.111, .177] \\

\hline
\end{tabular}
\end{table}

\section{Discussion}\label{sec5}

In this paper, we develop a model that employs Gaussian process priors to address practical violations of the positivity assumption when estimating causal effects from observational data. Unlike matching or trimming approaches, our method allows inference about the original target population. Further, unlike previous extrapolation methods, our approach does not require specifying arbitrary cut-offs in order to define nonoverlap regions. Importantly, our Gaussian process approach better reflects the greater uncertainty around estimated causal effects that is expected in areas of less covariate overlap. 

For complex outcome models containing nonlinearities and interactions, the GP model provided average treatment effect estimates with good performance. This result may be attributed to the nonparametric nature of the GP model and the centering of the GP prior of the prognostic component on a linear model. By pulling the prior in the direction of the data, more accurate estimates were obtained in data sparse areas. Further, the form of our model places a direct prior on the treatment effect, which may be beneficial for incorporating prior knowledge of the treatment effect and for subsequent interpretation--the form of the posterior for treatment effects is known. The GP model also provided more accurate and precise estimates of individual causal effects, which were most likely due to its accommodation of each subject's actual covariate values. For instance, for subjects in nonoverlap regions, we observed the GP model to be superior to the BART-only models in providing individual-level estimates that are close to the truth.

We emphasize the point that the type of positivity violation and the choice of approach that would be appropriate for addressing those violations depends on the clinical or study question of interest and the population of interest. If the objective is to account for all patient subgroups, that is, the set eligible for a particular treatment or intervention, then the population-level estimand is needed. There may be an extent of nonoverlap past which it would not be reasonable to obtain a population-level estimand--this may be a case when there are issues with study design or sampling, which should be dealt with before data are analyzed. Prior to fitting the GP model, standard exploratory analyses to assess covariate or propensity score overlap should be carried out and different definitions of overlap that have been proposed could be considered.\cite{zhu_2021} High degrees of nonoverlap may indicate that the current differences between the groups may result in invalid comparisons and that an alternative target population may be of more scientific or clinical interest.

In their invited discussion to Hahn et al. (2020), Papadogeorgou and Li suggest the Guassian process model may better address regions of poor overlap, where the GP model was shown to have larger measures of uncertainty.\cite{hahn_2020} In their example, they consider a single covariate and then fit separate outcome models for the treatment and control groups where the functions for each are given a GP prior. In our proposed approach, instead of placing priors on the outcome models for treated subjects and control subjects, respectively, we propose utilizing two GP priors in the same model for the prognostic and treatment effect components. With this choice of modeling, we allow data from both the treatment and control groups to influence the model fitting and thus more information is utilized in estimating the treatment effects. Our study furthers their illustrations by exploring the performance of the GP prior to address covariate nonoverlap in more complex data scenarios involving different degrees of nonoverlap and varying numbers of covariates. Further, by employing hyperpriors for the hyperparameters in the GP prior in the implementation of our model, results may be less sensitive to the particular prior specification.

One current limitation to our approach is the potential lack of scalability to very large studies due to computational challenges. The computational complexity is $\mathcal{O}(n^3)$ due to inversion of matrices that have dimensions equal to the sample size. Further, the long run time may be due to the number of parameters being sampled in our Markov Chain Monte Carlo; our algorithm estimates hyperparameters rather than fixing them at constant values.  In the covariance function, the same length-scale parameter is used for all the covariates. Modifying the kernel in the GP prior to allow for different length-scale parameters for different covariates may be beneficial especially when there is a large number of covariates. This may better capture the correlation between each covariate and outcome to reflect the relative importance of each variable and allow prior information regarding relevant confounders to be utilized. Moreover, here we have focused on continuous and binary outcomes. We are currently considering extensions of the GP approach to other outcomes that are common in clinical studies such as censored survival outcomes and longitudinal outcomes. In the longitudinal setting, addressing positivity violations may pose particular challenges due to time-dependent confounding.


\section*{Supporting information}

The following supporting information is available as part of the online article:

\noindent
\textbf{Figure S1.}
{Individual causal effect exploration when the continuous outcome is generated with $Y_1$ for the some nonoverlap setting.}

\noindent
\textbf{Figure S2.}
{Subject level mean and variability estimates for simulation setting c=0.}

\noindent
\textbf{Figure S3.}
{Subject level mean and variability estimates for simulation setting c=.70.}

\noindent
\textbf{Figure S4.}
{Different gamma distributions employed for the hyperpriors.}

\nocite{*}
\bibliographystyle{vancouver}
\bibliography{References}%

\begin{thebibliography}{10}

\bibitem{hernan_2006}
Hernán MA, Robins JM.
\newblock Estimating Causal Effects from Epidemiological Data.
\newblock Journal of epidemiology and community health. 2006-07;60(7):578-86.

\bibitem{westreich_2010}
Westreich D, Cole SR.
\newblock Invited Commentary: Positivity in Practice.
\newblock American journal of epidemiology. 2010-03-15;171(6):674-7; discussion
  678-81.

\bibitem{imbens_2015}
Imbens GW, Rubin DB.
\newblock Assessing Overlap in Covariate Distributions.
\newblock In: Causal Inference for Statistics, Social, and Biomedical Sciences:
  {{An}} Introduction. {Cambridge University Press}; 2015. p. 309-36.

\bibitem{damour_2020}
D’Amour A, Ding P, Feller A, Lei L, Sekhon J.
\newblock Overlap in Observational Studies with High-Dimensional Covariates.
\newblock Journal of Econometrics. 2020.

\bibitem{king_2006}
King G, Zeng L.
\newblock The {{Dangers}} of {{Extreme Counterfactuals}}.
\newblock Political Analysis. 2006;14(2):131-59.

\bibitem{petersen_2012}
Petersen ML, Porter KE, Gruber S, Wang Y, van~der Laan MJ.
\newblock Diagnosing and Responding to Violations in the Positivity Assumption.
\newblock Statistical Methods in Medical Research. 2012;21(1):31-54.

\bibitem{crump_2009}
Crump RK, Hotz VJ, Imbens GW, Mitnik OA.
\newblock Dealing with Limited Overlap in Estimation of Average Treatment
  Effects.
\newblock Biometrika. 2009;96(1):187-99.

\bibitem{Rosenbaum_2012}
Rosenbaum PR.
\newblock Optimal {{Matching}} of an {{Optimally Chosen Subset}} in
  {{Observational Studies}}.
\newblock Journal of Computational and Graphical Statistics. 2012;21(1):57-71.

\bibitem{visconti_2018}
Visconti G, Zubizarreta J.
\newblock Handling Limited Overlap in Observational Studies with Cardinality
  Matching.
\newblock Observational Studies. 2018;4:217-49.

\bibitem{hill_2013}
Hill J, Su YS.
\newblock Assessing Lack of Common Support in Causal Inference Using
  {{Bayesian}} Nonparametrics: {{Implications}} for Evaluating the Effect of
  Breastfeeding on Children’s Cognitive Outcomes.
\newblock The Annals of Applied Statistics. 2013-09;7(3):1386-420.

\bibitem{Ghosh_2018}
Ghosh D.
\newblock Relaxed Covariate Overlap and Margin-Based Causal Effect Estimation.
\newblock Statistics in Medicine. 2018;37(28):4252-65.

\bibitem{GhoshCruzCortes_2019}
Ghosh D, Cortes EC.
\newblock A Gaussian Process Framework for Overlap and Causal Effect Estimation
  with High-Dimensional Covariates.
\newblock Journal of Causal Inference. 2019;7(2):20180024.

\bibitem{li_2018}
Li F, Morgan KL, Zaslavsky AM.
\newblock Balancing {{Covariates}} via {{Propensity Score Weighting}}.
\newblock Journal of the American Statistical Association.
  2018;113(521):390-400.

\bibitem{li_2019}
Li F, Thomas LE, Li F.
\newblock Addressing {{Extreme Propensity Scores}} via the {{Overlap Weights}}.
\newblock American journal of epidemiology. 2019;188(1):250-7.

\bibitem{nethery_2019}
Nethery RC, Mealli F, Francesca D.
\newblock Estimating Population Average Causal Effects in the Presence of
  Non-Overlap: {{The}} Effect of Natural Gas Compressor Station Exposure on
  Cancer Mortality.
\newblock The Annals of Applied Statistics. 2019;13(2):1242-67.

\bibitem{rasmussen_2006}
Rasmussen CE, Williams CKI.
\newblock Gaussian {{Processes}} for {{Machine Learning}}.
\newblock The {{MIT Press}}. {Massachusetts Institute of Technology}; 2006.

\bibitem{neal_1998}
Neal RM.
\newblock Regression and Classification Using Gaussian Process Priors.
\newblock Bayesian Statistics 6. 1998.

\bibitem{rubin_2005}
Rubin DB.
\newblock Causal {{Inference Using Potential Outcomes}}.
\newblock Journal of the American Statistical Association.
  2005;100(469):322-31.

\bibitem{rosenbaum_1983}
Rosenbaum PR, Rubin DB.
\newblock The {{Central Role}} of the {{Propensity Score}} in {{Observational
  Studies}} for {{Causal Effects}}.
\newblock Biometrika. 1983;70(1):41-55.

\bibitem{rubin_2007}
Rubin DB.
\newblock The Design versus the Analysis of Observational Studies for Causal
  Effects: Parallels with the Design of Randomized Trials.
\newblock Statistics in Medicine. 2007;26(1):20-36.

\bibitem{hahn_2020}
Hahn PR, Murray JS, Carvalho CM.
\newblock Bayesian {{Regression Tree Models}} for {{Causal Inference}}:
  {{Regularization}}, {{Confounding}}, and {{Heterogeneous Effects}} (with
  {{Discussion}}).
\newblock Bayesian Anal. 2020-09;15(3):965-1056.

\bibitem{duvenaud_2014}
Duvenaud D.
\newblock Automatic model construction with Gaussian processes [Ph.D. thesis].
\newblock England: University of Cambridge; 2014.

\bibitem{genton_2002}
Genton MG.
\newblock Classes of Kernels for Machine Learning: A Statistics Perspective.
\newblock J Mach Learn Res. 2002 Marcj;2:299–312.

\bibitem{abramowitz_1965}
Abramowitz M, Stegun IA.
\newblock Handbook of Mathematical Functions with Formulas, Graphs, and
  Mathematical Tables.
\newblock {Dover Publications, Inc.}; 1965.

\bibitem{matern_1960}
Matern B.
\newblock Spatial Variation - Stochastic Models and Their Applications to Some
  Problems in Forest Survey Sampling Investigations.
\newblock No.~49 in Report of the {{Forest Research Institute}} of {{Sweden}}.
  Forest Research Institute; 1960.

\bibitem{uhlenbeck_1930}
Uhlenbeck GE, Ornstein LS.
\newblock On the Theory of the Brownian Motion.
\newblock Phys Rev. 1930 Sep;36:823-41.

\bibitem{craiu_2014}
Craiu RV, Rosenthal JS.
\newblock Bayesian {{Computation Via Markov Chain Monte Carlo}}.
\newblock Annual Review of Statistics and Its Application. 2014;1(1):179-201.

\bibitem{gelfand_2000}
Gelfand AE.
\newblock Gibbs {{Sampling}}.
\newblock Journal of the American Statistical Association. 2000;95(452):1300-4.

\bibitem{casella_1992}
Casella G, George EI.
\newblock Explaining the {{Gibbs Sampler}}.
\newblock The American Statistician. 1992;46(3):167-74.

\bibitem{gelman_2004}
Gelman A, Carlin JB, Stern HS, Rubin DB.
\newblock Bayesian Data Analysis.
\newblock 2nd ed. Chapman and Hall/CRC; 2004.

\bibitem{ellis_2018}
Ellis JA. A Practical Guide to MCMC Part 1: MCMC Basics; 2018.

\bibitem{meng_1999}
Meng XL, Van~Dyk DA.
\newblock Seeking {{Efficient Data Augmentation Schemes}} via {{Conditional}}
  and {{Marginal Augmentation}}.
\newblock Biometrika. 1999;86(2):301-20.

\bibitem{van_dyk_2001}
Van~Dyk DA, Meng XL.
\newblock The {{Art}} of {{Data Augmentation}}.
\newblock Journal of Computational and Graphical Statistics. 2001;10(1):1-50.

\bibitem{albert_1993}
Albert JH, Chib S.
\newblock Bayesian {{Analysis}} of {{Binary}} and {{Polychotomous Response
  Data}}.
\newblock Journal of the American Statistical Association.
  1993-06-01;88(422):669-79.

\bibitem{chipman_2010}
Chipman HA, George EI, McCulloch RE.
\newblock {{BART}}: {{Bayesian}} Additive Regression Trees.
\newblock The Annals of Applied Statistics. 2010;4(1):266-98.

\bibitem{r_2021}
{R Core Team}. R: A Language and Environment for Statistical Computing. Vienna,
  Austria; 2021.

\bibitem{connors_1996}
Connors AFJ, Speroff T, Dawson NV, Thomas C, Harrell FEJ, Wagner D, et~al.
\newblock The Effectiveness of Right Heart Catheterization in the Initial Care
  of Critically Ill Patients. {{SUPPORT Investigators}}.
\newblock JAMA. 1996;276(11):889-97.

\bibitem{zhu_2021}
Zhu Y, Hubbard RA, Chubak J, Roy J, Mitra N.
\newblock Core concepts in pharmacoepidemiology: Violations of the positivity
  assumption in the causal analysis of observational data: Consequences and
  statistical approaches.
\newblock Pharmacoepidemiology and Drug Safety. 2021;30(11):1471-85.

\bibitem{dehejia_2002}
Dehejia R, Wahba S.
\newblock Propensity {{Score Matching Methods For Non}}-{{Experimental Causal
  Studies}}.
\newblock The Review of Economics and Statistics. 2002;84:151-61.

\end{thebibliography}

\clearpage

\end{document}


\maketitle

\appendix

\section{Conditional distributions for $\mu$, $\beta$, and $\Delta$}
\label{app1}

For the choice of prior for the hyperparameter $\beta$ in the GP prior for $\mu$ is $p(\beta)\propto det(\sigma^2_{\beta}I_P)^{-1/2} \exp \left\{-\frac{1}{2} \beta^T (\sigma^2_{\beta}I_P)^{-1} \beta\right\}$, the conditional posterior distribution for $\beta$ is
\small
\begin{align*}
    & p(\beta|\mu, y) \propto p(y|\mu,\beta)p(\mu|\beta)p(\beta) \\
    & \propto \exp \left[-\frac{1}{2}\left\{y-(\mu+\Delta A)\right\}^T (\sigma^2I)^{-1}\left\{y-(\mu+\Delta A)\right\}\right] \exp \left\{-\frac{1}{2}(\mu-X\beta)^T K_{\mu}^{-1}(\mu-X\beta)\right\} \exp \left\{-\frac{1}{2}\beta^T (\sigma^2_{\beta}I_P)^{-1} \beta \right\} \\
    & \propto \exp \left[-\frac{1}{2}\left\{\mu^T K_{\mu}^{-1}\mu - 2\mu^T K_{\mu}^{-1} X\beta +\beta^T X^T K_{\mu}^{-1} X \beta + \beta^T (\sigma^2_{\beta}I_P)^{-1}\beta \right\}\right] \\
    & \propto \exp \left(-\frac{1}{2}\left[\beta^T \left\{X^T K_{\mu}^{-1}X+(\sigma^2_{\beta}I_P)^{-1}\right\} \beta - 2\beta^T X^T K_{\mu}^{-1} \mu \right]\right) \\
    & \propto \exp \left[-\frac{1}{2}\left(\beta^T \left\{X^T K_{\mu}^{-1}X+(\sigma^2_{\beta}I_P)^{-1}\right\} \beta - 2\beta^T\left\{X^T K_{\mu}^{-1}X+(\sigma^2_{\beta}I_P)^{-1}\right\}\left\{X^T K_{\mu}^{-1}X+(\sigma^2_{\beta}I_P)^{-1}\right\}^{-1}X^T K_{\mu}^{-1} \mu \right.\right. \\
    & \quad \quad \quad \quad \left. \left. + \left[\left\{X^T K_{\mu}^{-1}X+(\sigma^2_{\beta}I_P)^{-1}\right\}^{-1}X^T K_{\mu}^{-1}\mu\right]^T \left\{X^T K_{\mu}^{-1}X+(\sigma^2_{\beta}I_P)^{-1}\right\}\left[\left\{X^T K_{\mu}^{-1}X+(\sigma^2_{\beta}I_P)^{-1}\right\}^{-1}X^T K_{\mu}^{-1}\mu\right] \right)\right]\\
    & \propto \exp \left(-\frac{1}{2}\left[\beta-\left\{X^T K_{\mu}^{-1}X+(\sigma^2_{\beta}I_P)^{-1}\right\}^{-1}X^T K_{\mu}^{-1}\mu\right]^T 
    \left\{X^T K_{\mu}^{-1}X+(\sigma^2_{\beta}I_P)^{-1}\right\} \left[\beta-\left\{X^T K_{\mu}^{-1}X+(\sigma^2_{\beta}I_P)^{-1}\right\}^{-1}X^T K_{\mu}^{-1}\mu\right]\right)
\end{align*}
\normalsize
Thus, $\beta|\mu, y \sim MVN \left(\left[X^T K_{\mu}^{-1}X+(\sigma^2_{\beta}I_P)^{-1}\right]^{-1}X^T K_{\mu}^{-1}\mu, \left[X^T K_{\mu}^{-1}X+(\sigma^2_{\beta}I_P)^{-1}\right]^{-1}\right)$.

The prior for $\mu$ is $p(\mu) \propto det(K_{\mu})^{-\frac{1}{2}} \exp \left[-\frac{1}{2}(\mu-X\beta)^TK_{\mu}^{-1}(\mu-X\beta)\right]$. 

The prior for $\Delta$ is $p(\Delta) \propto det(K_{\Delta})^{-\frac{1}{2}} \exp\left[-\frac{1}{2}\Delta^T K_{\Delta}^{-1} \Delta \right]$.

Assuming prior independence of $\mu$ and $\Delta$, $p(\mu,\Delta) = p(\mu)p(\Delta)$, the joint posterior is 
\begin{align*}
    p(\mu,\Delta|y) &\propto p(y|\mu,\Delta)p(\mu,\Delta) \\
    &\propto p(y|\mu, \Delta) p(\mu) p(\Delta)
\end{align*}

The posterior distributions for $\mu$ and $\Delta$ are obtained as follows.

The posterior for $\mu|\Delta, y$ is given by
\small
\begin{align*}
    & p(\mu|\Delta, y) \propto p(y|\mu, \Delta) p(\mu) \\
    &\propto det(\sigma^2I)^{-\frac{1}{2}} \exp \left[-\frac{1}{2}\left\{y-(\mu+\Delta A)\right\}^T(\sigma^2I)^{-1} \left\{y-(\mu+\Delta A)\right\}\right] det(K_{\mu})^{-\frac{1}{2}} \exp \left[-\frac{1}{2}(\mu-X\beta)^TK_{\mu}^{-1}(\mu-X\beta)\right] \\
    &\propto \exp \left(-\frac{1}{2}\left[\left\{y-(\mu+\Delta A)\right\}^T(\sigma^2I)^{-1}\left\{y-(\mu+\Delta A)\right\}+(\mu-X\beta)^T K_{\mu}^{-1} (\mu-X\beta)\right]\right) \\
    &\propto \exp \left[-\frac{1}{2} \left\{ y^T(\sigma^2I)^{-1}y-2y^T(\sigma^2I)^{-1}(\mu+\Delta A) + (\mu+\Delta A)(\sigma^2I)^{-1}(\mu+\Delta A) + \mu^T K_{\mu}^{-1} \mu -2\mu^TK_{\mu}^{-1}X\beta + (X\beta)^TK_{\mu}^{-1} X\beta \right\}\right] \\
    &\propto \exp \left[-\frac{1}{2}\left\{-2\mu^T(\sigma^2I)^{-1}y + \mu^T(\sigma^2I)^{-1} \mu + 2\mu^T(\sigma^2I)^{-1}\Delta A +\mu^T K_{\mu}^{-1} \mu -2\mu^T K_{\mu}^{-1}X\beta \right\}\right]\\
    &\propto \exp \left(-\frac{1}{2}\left[\mu^T\left\{K_{\mu}^{-1}+(\sigma^2I)^{-1}\right\} \mu-2\mu^T\left\{(\sigma^2I)^{-1}y-(\sigma^2I)^{-1}\Delta A+K_{\mu}^{-1}X\beta \right\}\right]\right) \\
    &\propto \exp \left[-\frac{1}{2}\left(\mu^T \left\{K_{\mu}^{-1}+(\sigma^2I)^{-1} \right\} \mu -2\mu^T \left\{K_{\mu}^{-1}+(\sigma^2I)^{-1}\right\} \left\{K_{\mu}^{-1}+(\sigma^2I)^{-1}\right\}^{-1}\left\{(\sigma^2I)^{-1}(y-\Delta A)+K_{\mu}^{-1}X\beta \right\} \right. \right. \\
    & \quad \quad \quad \quad + \left[\left\{K_{\mu}^{-1}+(\sigma^2I)^{-1}\right\}^{-1}\left\{(\sigma^2I)^{-1}(y-\Delta A)+K_{\mu}^{-1}X\beta\right\}\right]^T \left\{K_{\mu}^{-1}+(\sigma^2I)^{-1}\right\} \\
    & \quad \quad \quad \quad \times \left. \left. \left\{K_{\mu}^{-1}+(\sigma^2I)^{-1}\right\}^{-1}\left\{(\sigma^2I)^{-1}(y-\Delta A)+K_{\mu}^{-1}X\beta\right\} \right) \right] \\
    &\propto \exp \left(-\frac{1}{2}\left[\mu-\left\{K_{\mu}^{-1}+(\sigma^2I)^{-1}\right\}^{-1}\left\{(\sigma^2I)^{-1}(y-\Delta A)+K_{\mu}^{-1}X\beta\right\}\right]^T \left\{K_{\mu}^{-1}+(\sigma^2I)^{-1}\right\} \right. \\
    &\quad \quad \quad \quad \left. \left[\mu-\left\{K_{\mu}^{-1}+(\sigma^2I)^{-1}\right\}^{-1}\left\{(\sigma^2I)^{-1}(y-\Delta A)+K_{\mu}^{-1}X\beta\right\}\right] \right)
\end{align*}
\normalsize
Thus, $\mu|\Delta, y \sim MVN \left( \left[K_{\mu}^{-1}+(\sigma^2I)^{-1}\right]^{-1}\left[(\sigma^2I)^{-1}(y-\Delta A)+K_{\mu}^{-1}X\beta \right], \left[K_{\mu}^{-1}+(\sigma^2I)^{-1}\right]^{-1}\right)$.

The posterior for $\Delta|\mu, y$ is given by
\small
\begin{align*}
     &p(\Delta|\mu, y) \propto p(y|\mu, \Delta) p(\Delta) \\
     &\propto det(\sigma^2I)^{-\frac{1}{2}} \exp \left[\left\{y-(\mu+\Delta A)\right\}^T(\sigma^2I)^{-1} \left\{y-(\mu+\Delta A)\right\}\right] det(K_{\Delta})^{-\frac{1}{2}} \exp \left(-\frac{1}{2}\Delta^T K_{\Delta}^{-1} \Delta \right) \\
     &\propto \exp \left[-\frac{1}{2}\left\{y^T(\sigma^2I)^{-1}y-2y^T(\sigma^2I)^{-1}(\mu+\Delta A) + (\mu+\Delta A)^T (\sigma^2I)^{-1} (\mu+\Delta A) + \Delta^T K_{\Delta}^{-1} \Delta \right\}\right] \\
     &\propto \exp \left[-\frac{1}{2}\left\{-2(\Delta A)^T(\sigma^2I)^{-1}y + \mu^T(\sigma^2I)^{-1}\mu +2(\Delta A)^T(\sigma^2I)^{-1} \mu + (\Delta A)^T (\sigma^2I)^{-1} \Delta A + \Delta^T K_{\Delta}^{-1} \Delta \right\}\right] \\
     &\propto \exp \left(-\frac{1}{2}\left[\Delta^T \left\{A^T\odot (\sigma^2I)^{-1} \odot A +  K_{\Delta}^{-1}\right\} \Delta -2\Delta^T\left\{A^T \odot (\sigma^2I)^{-1}(y-\mu)\right\}\right]\right) \\
     &\propto \exp \left(-\frac{1}{2}\left[\Delta^T \left\{K_{\Delta}^{-1}+A^T\odot (\sigma^2I)^{-1} \odot A\right\}\Delta -2\Delta^T \left\{K_{\Delta}^{-1}+A^T\odot (\sigma^2I)^{-1} \odot A\right\} \left\{K_{\Delta}^{-1}+A^T\odot (\sigma^2I)^{-1} \odot A\right\}^{-1} \right. \right. \\
     &\quad \quad \quad \quad \left. \left. \left\{A^T \odot (\sigma^2I)^{-1}(y-\mu)\right\} + (y-\mu)^T \left[\left\{K_{\Delta}^{-1}+A^T\odot (\sigma^2I)^{-1} \odot A\right\}^{-1} A^T \odot (\sigma^2I)^{-1}\right]^T \right.\right. \\
     &\quad \quad \quad \quad \left. \left. \left\{K_{\Delta}^{-1}+A^T\odot (\sigma^2I)^{-1} \odot A\right\} \left\{K_{\Delta}^{-1}+A^T\odot (\sigma^2I)^{-1} \odot A\right\}^{-1} A^T \odot (\sigma^2I)^{-1} (y-\mu) \right]\right) \\
     &\propto \exp \left(-\frac{1}{2}\left[\Delta - \left\{K_{\Delta}^{-1}+A^T\odot (\sigma^2I)^{-1} \odot A\right\}^{-1} A^T \odot (\sigma^2I)^{-1} (y-\mu) \right]^T \left\{K_{\Delta}^{-1}+A^T\odot (\sigma^2I)^{-1} \odot A\right\} \right. \\
     &\quad \quad \quad \quad \left. \left[\Delta - \left\{K_{\Delta}^{-1}+A^T\odot (\sigma^2I)^{-1} \odot A\right\}^{-1} A^T \odot (\sigma^2I)^{-1} (y-\mu) \right] \right)
\end{align*}
\normalsize
Thus, $\Delta|\mu, y \sim MVN\left(\left[K_{\Delta}^{-1}+A^T\odot (\sigma^2I)^{-1} \odot A \right]^{-1} A^T \odot (\sigma^2I)^{-1} (y-\mu), \left[K_{\Delta}^{-1}+A^T\odot (\sigma^2I)^{-1} \odot A\right]^{-1}\right)$.

\section{Steps of the Metropolis-within-Gibbs Algorithm}
\label{app2}
In this section, we present the steps of the algorithm for obtaining posterior sampling of the parameters and hyperparameters. Let $l(\mu, \beta, l_{\mu}, \eta_{\mu}, \Delta, l_{\Delta}, \eta_{\Delta}, \sigma^2) = \log p(\mu, \beta, l_{\mu}, \eta_{\mu}, \Delta, l_{\Delta}, \eta_{\Delta}, \sigma^2|Y)$ denote the log posterior distribution, and let $q(\text{ } \cdot \text{ };m, s^2)$ be the density of the proposal distribution with mean $m$ and variance $s^2$. We start the chains with initial values
\begin{equation*}
    \mu^{(0)}, \beta^{(0)}, l_{\mu}^{(0)}, \eta_{\mu}^{(0)}, \Delta^{(0)}, l_{\Delta}^{(0)}, \eta_{\Delta}^{(0)}, \sigma^{2(0)}
\end{equation*}

At iteration $j$,

\begin{enumerate}
     \item Draw $l_{\mu}^*$ from the proposal distribution--truncated normal distribution centered at $l_{\mu}^{(j-1)}$ with variance $\tau_{l_\mu}^2$ and bounded below at 0:
    \begin{equation*}
        l_{\mu}^* \sim TN(l_{\mu}^{(j-1)}, \tau_{l_\mu}^2; lower=0)
    \end{equation*}
    \begin{align*}
        \log r_{l_\mu} &= l(l_{\mu}^*, \eta_{\mu}^{(j-1)},  \beta^{(j-1)}, \mu^{(j-1)}, l_{\Delta}^{(j-1)}, \eta_{\Delta}^{(j-1)}, \Delta^{(j-1)}, \sigma^{2(j-1)}) \\
        & \quad \quad - l(l_{\mu}^{(j-1)}, \eta_{\mu}^{(j-1)}, \beta^{(j-1)}, \mu^{(j-1)}, l_{\Delta}^{(j-1)}, \eta_{\Delta}^{(j-1)}, \Delta^{(j-1)}, \sigma^{2(j-1)}) \\
        & \quad\quad +\log[q(l_{\mu}^{(j-1)};l_{\mu}^*, \tau_{l_{\mu}}^2)]-\log[q(l_{\mu}^*;l_{\mu}^{(j-1)},\tau_{l_{\mu}}^2)] \\
    \end{align*}
    We then draw a random $U \sim Unif(0,1)$ and set
        \begin{equation*}
            l_{\mu}^{(j)}=
        \begin{cases}
            l_{\mu}^*, & \text{if}\ \log U \leq \log r_{l_{\mu}} \\
            l_{\mu}^{(j-1)}, & \text{otherwise}
        \end{cases}
  \end{equation*}
  \item Draw $\eta_{\mu}^*$ from the proposal distribution--truncated normal distribution centered at $\eta_{\mu}^{(j-1)}$ with variance $\tau_{\eta_\mu}^2$ and bounded below at 0:
    \begin{equation*}
        \eta_{\mu}^* \sim TN(\eta_{\mu}^{(j-1)}, \tau_{\eta_\mu}^2; lower=0)
    \end{equation*}
    \begin{align*}
        \log r_{\eta_\mu} &= l(l_{\mu}^{(j)}, \eta_{\mu}^*, \beta^{(j-1)}, \mu^{(j-1)}, l_{\Delta}^{(j-1)}, \eta_{\Delta}^{(j-1)}, \Delta^{(j-1)}, \sigma^{2(j-1)}) \\
        & \quad \quad - l(l_{\mu}^{(j)}, \eta_{\mu}^{(j-1)}, \beta^{(j-1)}, \mu^{(j-1)}, l_{\Delta}^{(j-1)}, \eta_{\Delta}^{(j-1)}, \Delta^{(j-1)}, \sigma^{2(j-1)}) \\
        & \quad\quad +\log[q(\eta_{\mu}^{(j-1)};\eta_{\mu}^*, \tau_{\eta_{\mu}}^2)]-\log[q(\eta_{\mu}^*;\eta_{\mu}^{(j-1)},\tau_{\eta_{\mu}}^2)] \\
    \end{align*}
    We then draw a random $U \sim Unif(0,1)$ and set
        \begin{equation*}
            \eta_{\mu}^{(j)}=
        \begin{cases}
            \eta_{\mu}^*, & \text{if}\ \log U \leq \log r_{\eta_{\mu}} \\
            \eta_{\mu}^{(j-1)}, & \text{otherwise}
        \end{cases}
  \end{equation*}
  
  \item Draw $\beta^{(j)}$ from
  \begin{align*}
    MVN & \left((X^T K_{\mu}(l_{\mu}^{(j)}, \eta_{\mu}^{(j)})^{-1}X+(\sigma^2_{\beta}I_P)^{-1})^{-1}X^T K_{\mu}(l_{\mu}^{(j)}, \eta_{\mu}^{(j)})^{-1}\mu^{(j-1)}, \right. \\
    & \quad \quad \left. [X^T K_{\mu}(l_{\mu}^{(j)}, \eta_{\mu}^{(j)})^{-1}X+(\sigma^2_{\beta}I_P)^{-1}]^{-1}\right)
\end{align*}
  \item Draw $\mu^{(j)}$ from \begin{align*}
      MVN(& [K_{\mu}(l_{\mu}^{(j)}, \eta_{\mu}^{(j)})^{-1}+(\sigma^{2(j-1)}I)^{-1}]^{-1}[(\sigma^{2(j-1)}I)^{-1}(y-\Delta^{(j-1)} A) + K_{\mu}^{-1}X\beta^{(j)}], \\
      & \quad \quad [K_{\mu}(l_{\mu}^{(j)}, \eta_{\mu}^{(j)})^{-1}+(\sigma^{2(j-1)}I)^{-1}]^{-1})
  \end{align*}
  \item Draw $l_{\Delta}^*$ from its proposal distribution:
    \begin{equation*}
        l_{\Delta}^* \sim TN(l_{\Delta}^{(j-1)}, \tau_{l_\Delta}^2; lower=0)
    \end{equation*}
    \begin{align*}
        \log r_{l_\Delta} &= l(l_{\mu}^{(j)}, \eta_{\mu}^{(j)}, \beta^{(j)}, \mu^{(j)}, l_{\Delta}^*, \eta_{\Delta}^{(j-1)}, \Delta^{(j-1)}, \sigma^{2(j-1)}) \\
        & \quad \quad - l(l_{\mu}^{(j)}, \eta_{\mu}^{(j)}, \beta^{(j)}, \mu^{(j)}, l_{\Delta}^{(j-1)}, \eta_{\Delta}^{(j-1)}, \Delta^{(j-1)}, \sigma^{2(j-1)}) \\
        & \quad\quad +\log[q(l_{\Delta}^{(j-1)};l_{\Delta}^*, \tau_{l_{\Delta}}^2)]-\log[q(l_{\Delta}^*;l_{\Delta}^{(j-1)},\tau_{l_{\Delta}}^2)] \\
    \end{align*}
    We then draw a random $U \sim Unif(0,1)$ and set
        \begin{equation*}
            l_{\Delta}^{(j)}=
        \begin{cases}
            l_{\Delta}^*, & \text{if}\ \log U \leq \log r_{l_{\Delta}} \\
            l_{\Delta}^{(j-1)}, & \text{otherwise}
        \end{cases}
  \end{equation*}
  \item Draw $\eta_{\Delta}^*$ from its proposal distribution:
    \begin{equation*}
        \eta_{\Delta}^* \sim TN(\eta_{\Delta}^{(j-1)}, \tau_{\eta_\Delta}^2; lower=0)
    \end{equation*}
    \begin{align*}
        \log r_{\eta_\Delta} &= l(l_{\mu}^{(j)}, \eta_{\mu}^{(j)}, \beta^{(j)}, \mu^{(j)}, l_{\Delta}^{(j)}, \eta_{\Delta}^*, \Delta^{(j-1)}, \sigma^{2(j-1)}) \\
        & \quad \quad - l(l_{\mu}^{(j)}, \eta_{\mu}^{(j)}, \beta^{(j)}, \mu^{(j)}, l_{\Delta}^{(j)}, \eta_{\Delta}^{(j-1)}, \Delta^{(j-1)}, \sigma^{2(j-1)}) \\
        & \quad\quad +\log[q(\eta_{\Delta}^{(j-1)};\eta_{\Delta}^*, \tau_{\eta_{\Delta}}^2)]-\log[q(\eta_{\Delta}^*;\eta_{\Delta}^{(j-1)},\tau_{\eta_{\Delta}}^2)] \\
    \end{align*}
    We then draw a random $U \sim Unif(0,1)$ and set
        \begin{equation*}
            \eta_{\Delta}^{(j)}=
        \begin{cases}
            \eta_{\Delta}^*, & \text{if}\ \log U \leq \log r_{\eta_{\Delta}} \\
            \eta_{\Delta}^{(j-1)}, & \text{otherwise}
        \end{cases}
  \end{equation*}
  \item Draw $\Delta^{(j)}$ from
  \begin{equation*}
      MVN([K_{\Delta}^{-1}+A^T\odot (\sigma^2I)^{-1} \odot A]^{-1} A^T \odot (\sigma^2I)^{-1} (y-\mu), [K_{\Delta}^{-1}+A^T\odot (\sigma^2I)^{-1} \odot A]^{-1})
  \end{equation*}
  \item Draw $\sigma^{2*}$ from its proposal distribution:
    \begin{equation*}
        \sigma^{2*} \sim TN(\sigma^{2(j-1)}, \tau_{\sigma^2}^2; lower=0)
    \end{equation*}
    \begin{align*}
        \log r_{\sigma^2} &= l(l_{\mu}^{(j)}, \eta_{\mu}^{(j)}, \beta^{(j)}, \mu^{(j)}, l_{\Delta}^{(j)}, \eta_{\Delta}^{(j)}, \Delta^{(j)}, \sigma^{2*}) \\
        & \quad \quad - l(l_{\mu}^{(j)}, \eta_{\mu}^{(j)}, \beta^{(j)}, \mu^{(j)}, l_{\Delta}^{(j)}, \eta_{\Delta}^{(j)}, \Delta^{(j)}, \sigma^{2(j-1)}) \\
        & \quad\quad +\log[q(\sigma^{2(j-1)};\sigma^{2*}, \tau_{\sigma^2}^2)]-\log[q(\sigma^{2*};\sigma^{2(j-1)},\tau_{\sigma^2}^2)] \\
    \end{align*}
    We then draw a random $U \sim Unif(0,1)$ and set
        \begin{equation*}
            \sigma^{2(j)}=
        \begin{cases}
            \sigma^{2*}, & \text{if}\ \log U \leq \log r_{\sigma^2} \\
            \sigma^{2(j-1)}, & \text{otherwise}
        \end{cases}
  \end{equation*}
\end{enumerate}
This continues until the number the posterior draws after thinning and burn-ins (say, $J$) is reached.

\section{Results for Continuous Outcome Scenario Involving Nonlinearity and Treatment Heterogeneity}
\label{app3}

\begin{table}[H]
\centering
\caption{Effect estimates for nonoverlap scenarios involving a nonlinear response surface and treatment heterogeneity. The true ATE for the some nonoverlap and substantial nonoverlap settings are .950 and .564, respectively.}
\begin{tabular}{c|c|ccccccc}
\hline
                                                                                  & Method          & ATE   & Bias   & \% Bias  & $\overline{SD}$ & SE   & MSE   & Coverage \\ \hline
\multirow{6}{*}{\begin{tabular}[c]{@{}c@{}}Some \\ nonoverlap\end{tabular}}       & GP              & .849  & -.101  & -10.612  & .119            & .225 & .061  & .657     \\
                                                                                  & BCF             & .814  & -.136  & -14.298  & .140            & .251 & .082  & .658     \\
                                                                                  & BART-Stratified & .758  & -.192  & -20.240  & .118            & .237 & .093  & .535     \\
                                                                                  & BART-Single     & .658  & -.292  & -30.695  & .205            & .252 & .149  & .667     \\
                                                                                  & BART+SPL        & .580  & -.370  & -38.918  & .254            & .285 & .218  & .689     \\
                                                                                  & Probit model    & -.049 & -.999  & -105.120 & .315            & .279 & 1.075 & .082     \\ \hline
\multirow{6}{*}{\begin{tabular}[c]{@{}c@{}}Substantial\\ nonoverlap\end{tabular}} & GP              & .411  & -.153  & -27.154  & .149            & .242 & .082  & .692     \\
                                                                                  & BCF             & .349  & -.215  & -38.079  & .168            & .256 & .111  & .663     \\
                                                                                  & BART-Stratified & .344  & -.219  & -38.898  & .148            & .247 & .109  & .588     \\
                                                                                  & BART-Single     & .190  & -.374  & -66.373  & .218            & .258 & .207  & .589     \\
                                                                                  & BART+SPL        & .013  & -.550  & -97.639  & .386            & .357 & .430  & .707     \\
                                                                                  & Probit model    & -.562 & -1.125 & -199.632 & .338            & .319 & 1.368 & .088     \\ \hline
\end{tabular}
\end{table}

\begin{figure}[H]
    \centering
    \includegraphics[width=15cm]{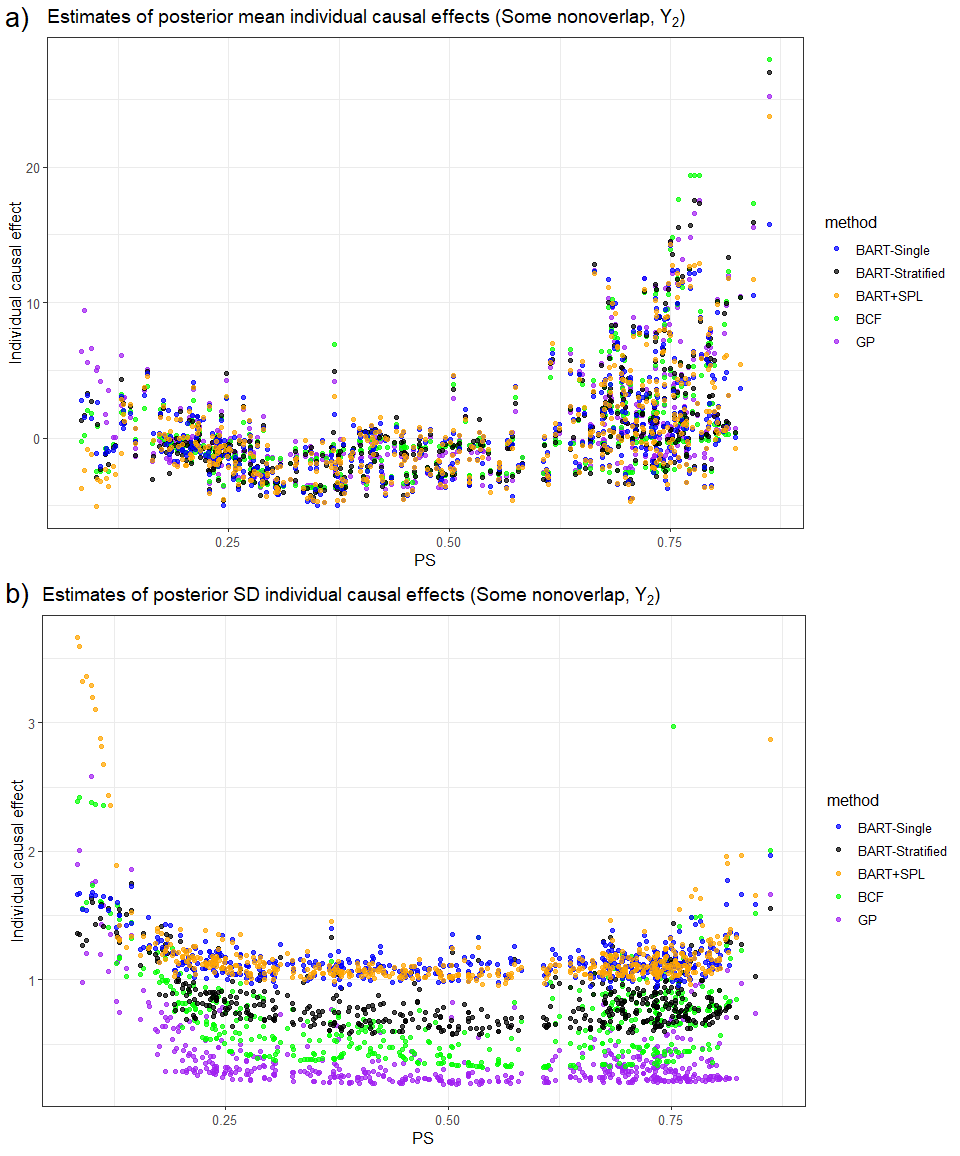}
    \caption{Individual causal effect exploration when the continuous outcome is generated with $Y_2$ for the some nonoverlap setting.}
    \label{fig:1NL-ICE}
\end{figure}

\begin{figure}[H]
    \centering
    \includegraphics[width=15cm]{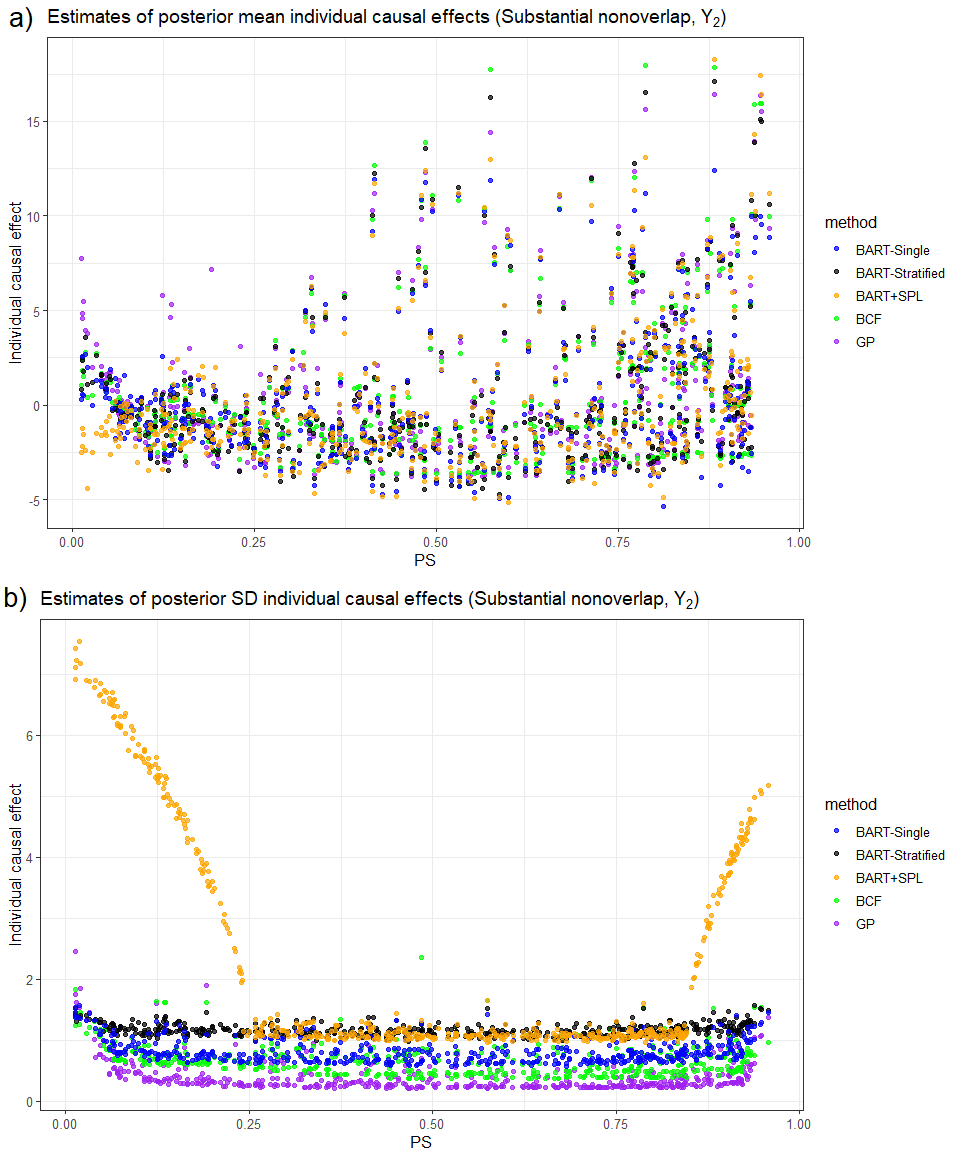}
    \caption{Individual causal effect exploration when the continuous outcome is generated with $Y_2$ for the substantial nonoverlap setting.}
    \label{fig:2NL-ICE}
\end{figure}

Figure \ref{fig:1NL-ICE} and \ref{fig:2NL-ICE} provide further information regarding individual causal effects estimated by each method. The GP model is able to capture the larger average treatment effects for subjects with estimated propensity scores near 0, which helps to pull up its estimates of the ATE. Furthermore, for these subjects, estimates of posterior standard deviations obtained from the GP model are larger than those from both BART models. This suggests that the continuous nature of the GP model better allows larger distances to be translated into greater estimates of uncertainty as compared to the BART models. On the other hand, BART+SPL's variation inflation factor greatly overestimates the corresponding uncertainty, which results in inconsequential knowledge about the treatment effects for subjects in nonoverlap areas.

\newpage 

\newpage

\section{Simulation Results Looking at Gaussian Process Prior Specifications}
\label{app4}
\begin{table}[H]
\centering
\caption{Simulation results for two representative simulation settings based on different choices of hyperpriors in the Gaussian process priors.}
\label{tab:hyperprior}
\begin{tabular}{cc|ccccccc}
\hline
                                                                                                    &            & ATE   & Bias  & \% Bias & $\overline{SD}$ & SE   & MSE                    & Coverage \\ \hline
\multirow{2}{*}{\begin{tabular}[c]{@{}c@{}}Some nonoverlap,\\ Linear response surface\end{tabular}} & Gamma(2,1) & 1.968 & -.032 & -1.603  & .102            & .010 & .011                   & .948     \\
                                                                                                    & Gamma(4,.5) & 1.993 & -.007 & -367    & .101            & .097 & .009                   & .960     \\ \hline
\multirow{2}{*}{\begin{tabular}[c]{@{}c@{}}Nethery et al. setting\\ with c=.35\end{tabular}}        & Gamma(2,1) & -.190 & -.002 & -1.583  & .030            & .057 & $2.909 \times 10^{-4}$ & .998     \\
                                                                                                    & Gamma(4,.5) & -.185 & .001  & .641    & .029            & .058 & $2.344\times 10^{-4}$  & 1.000    \\ \hline
\end{tabular}
\end{table}

\begin{table}[H]
\caption{Simulation results from two representative simulation settings based on different choices in the covariance function in the Gaussian process priors.}
\label{tab:kernel}
\small
\begin{tabular}{cc|ccccccc}
\hline
                                                                                                    &                         & ATE   & Bias  & \% Bias & $\overline{SD}$ & SE   & MSE                     & Coverage \\ \hline
\multirow{4}{*}{\begin{tabular}[c]{@{}c@{}}Some nonoverlap,\\ Linear response surface\end{tabular}} & Squared exponential     & 1.968 & -.032 & -1.603  & .102            & .010 & .011                    & .948     \\
                                                                                                    & Rational quadratic      & 1.971 & -.029 & -1.445  & .104            & .098 & .011                    & .954     \\
                                                                                                    & Mat\'ern & 1.965 & -.035 & -1.737  & .104            & .099 & .011                    & .956     \\
                                                                                                    & Ornstein–Uhlenbeck      & 1.951 & -.049 & -2.430  & .104            & .099 & .012                    & .944     \\ \hline
\multirow{4}{*}{\begin{tabular}[c]{@{}c@{}}Nethery et al. setting\\ with c=.35\end{tabular}}        & Squared exponential     & -.190 & -.002 & -1.583  & .030            & .057 & $2.909 \times 10^{-4}$  & .998     \\
                                                                                                    & Rational quadratic      & -.186 & .001  & .505    & .028            & .057 & $1.119 \times 10^{-4}$  & 1.000    \\
                                                                                                    & Mat\'ern & -.183 & .002  & .908    & .027            & .056 & $7.251 \times 10^ {-5}$ & 1.000    \\
                                                                                                    & Ornstein–Uhlenbeck      & -.183 & .004  & 2.034   & .027            & .058 & $1.691 \times 10^{-4}$  & 1.000    \\ \hline
\end{tabular}
\end{table}

\newpage
\section{Results for Binary Outcomes}
\label{app5}

\begin{table}[H]
\centering
\caption{Effect estimates from each method for nonoverlap scenarios involving a outcome model ($Y_{1B}$) that is linear on the probit scale. The true ATE is .280 for the some nonoverlap setting and .283 for the substantial nonoverlap setting.}
\begin{tabular}{c|c|ccccccc}
\hline
                                                                                  & Method          & ATE  & Bias                    & \% Bias & $\overline{SD}$ & SE   & MSE  & Coverage \\ \hline
\multirow{6}{*}{\begin{tabular}[c]{@{}c@{}}Some \\ nonoverlap\end{tabular}}       & GP              & .269 & -.010                   & -3.602  & .024            & .026 & .001 & .902     \\
                                                                                  & BCF             & .300 & .020                    & 7.242   & .035            & .028 & .001 & .957     \\
                                                                                  & BART-Stratified & .249 & -.031                   & -11.008 & .030            & .028 & .002 & .833     \\
                                                                                  & BART-Single     & .270 & -.010                   & -3.535  & .028            & .026 & .001 & .947     \\
                                                                                  & BART+SPL        & .276 & -.004                   & -1.393  & .033            & .033 & .001 & .952     \\
                                                                                  & Probit model    & .279 & $-4.652 \times 10^{-4}$ & -.166   & .011            & .026 & .001 & .614     \\ \hline
\multirow{6}{*}{\begin{tabular}[c]{@{}c@{}}Substantial\\ nonoverlap\end{tabular}} & GP              & .274 & -.009                   & -3.044  & .030            & .032 & .001 & .916     \\
                                                                                  & BCF             & .279 & -.004                   & -1.313  & .041            & .036 & .001 & .976     \\
                                                                                  & BART-Stratified & .267 & -.016                   & -5.590  & .039            & .034 & .001 & .964     \\
                                                                                  & BART-Single     & .271 & -.012                   & -4.131  & .036            & .031 & .001 & .970     \\
                                                                                  & BART+SPL        & .280 & -.003                   & -.894   & .054            & .043 & .002 & .984     \\
                                                                                  & Probit model    & .283 & $-3.862 \times 10^{-4}$ & -.136   & .011            & .031 & .001 & .517     \\ \hline
\end{tabular}
\end{table}

\begin{table}[H]
\centering
\caption{Effect estimates from each method for nonoverlap scenarios involving a outcome model ($Y_{2B}$) that is nonlinear and involves interactions on the probit scale. The true ATE is -.146 for the some nonoverlap setting and -.202 for the substantial nonoverlap setting.}

\begin{tabular}{c|c|ccccccc}
\hline
                                                                                  & Method          & ATE   & Bias  & \% Bias & $\overline{SD}$ & SE   & MSE  & Coverage \\ \hline
\multirow{6}{*}{\begin{tabular}[c]{@{}c@{}}Some \\ nonoverlap\end{tabular}}       & GP              & -.157 & -.011 & -7.785  & .029            & .036 & .001 & .863     \\
                                                                                  & BCF             & -.148 & -.002 & -1.297  & .031            & .038 & .001 & .882     \\
                                                                                  & BART-Stratified & -.197 & -.051 & -35.129 & .037            & .038 & .004 & .697     \\
                                                                                  & BART-Single     & -.199 & -.053 & -36.505 & .044            & .040 & .004 & .790     \\
                                                                                  & BART+SPL        & -.213 & -.067 & -46.190 & .050            & .051 & .007 & .736     \\
                                                                                  & Probit model    & -.259 & -.113 & -77.437 & .001            & .048 & .015 & .001     \\ \hline
\multirow{6}{*}{\begin{tabular}[c]{@{}c@{}}Substantial\\ nonoverlap\end{tabular}} & GP              & -.220 & -.017 & -8.456  & .039            & .045 & .002 & .893     \\
                                                                                  & BCF             & -.212 & -.010 & -4.696  & .038            & .047 & .002 & .877     \\
                                                                                  & BART-Stratified & -.254 & -.052 & -25.639 & .043            & .043 & .005 & .781     \\
                                                                                  & BART-Single     & -.268 & -.065 & -32.275 & .048            & .045 & .006 & .750     \\
                                                                                  & BART+SPL        & -.297 & -.094 & -46.567 & .066            & .057 & .012 & .729     \\
                                                                                  & Probit model    & -.332 & -.130 & -64.012 & .001            & .052 & .020 & 0        \\ \hline
\end{tabular}
\end{table}

\section{\large Patient Characteristics from the Right Heart Catheterization Study}
\label{app6}
\vspace{-4ex}
\begin{table}[H]
\centering
\caption{\small Characteristics of patients who received a right heart catheterization and those who did not. Continuous variables are represented by mean (SD); categorical variables are represented by n (\%).}
\label{tab:table1}
\resizebox{14cm}{!}{
\begin{tabular}{cc|ccc}
\hline
                                   &                                     & No RHC               & RHC                  & p-value              \\ \hline
n                                  &                                     & 480                  & 137                  &                      \\ \hline
Age                                &                                     & 61.64 (18.02)        & 58.04 (16.33)        & .036                 \\ \hline
Race                               &                                     &                      &                      & .364                 \\
                                   & white                               & 361 (75.2)           & 97 (70.8)            &                      \\
                                   & black                               & 92 (19.2)            & 28 (20.4)            &                      \\
                                   & other                               & 27 (5.6)             & 12 (8.8)             &                      \\ \hline
Education (years)                  &                                     & 11.60 (2.89)         & 12.03 (2.43)         & .109                 \\ \hline
Income (\$)                        &                                     &                      &                      & .004                 \\
                                   & \textless{}11k                      & 291 (60.6)           & 71 (51.8)            &                      \\
                                   & 11-25k                              & 105 (21.9)           & 23 (16.8)            &                      \\
                                   & 25-50k                              & 58 (12.1)            & 32 (23.4)            &                      \\
                                   & \textgreater{}50k                   & 26 (5.4)             & 11 (8.0)             &                      \\ \hline
Medical insurance                  &                                     &                      &                      & .146                 \\
                                   & Private                             & 118 (24.6)           & 50 (36.5)            &                      \\
                                   & Medicare                            & 144 (30.0)           & 35 (25.5)            &                      \\
                                   & Medicaid                            & 74 (15.4)            & 16 (11.7)            &                      \\
                                   & Private \& Medicare                 & 83 (17.3)            & 21 (15.3)            &                      \\
                                   & Medicare \& Medicaid                & 38 (7.9)             & 8 (5.8)              &                      \\
                                   & No insurance                        & 23 (4.8)             & 7 (5.1)              &                      \\ \hline
Primary disease category           &                                     &                      &                      & \textless{}.001      \\
                                   & ARF                                 & 201 (41.9)           & 50 (36.5)            &                      \\
                                   & CHF                                 & 70 (14.6)            & 25 (18.2)            &                      \\
                                   & Cirrhosis                           & 26 (5.4)             & 2 (1.5)              &                      \\
                                   & Colon Cancer                        & 1 (0.2)              & 0 (0.0)              &                      \\
                                   & Coma                                & 1 (0.2)              & 0 (0.0)              &                      \\
                                   & COPD                                & 83 (17.3)            & 6 (4.4)              &                      \\
                                   & Lung Cancer                         & 3 (0.6)              & 1 (0.7)              &                      \\
                                   & MOSF with Malignancy                & 35 (7.3)             & 9 (6.6)              &                      \\
                                   & MOSF with Sepsis                    & 60 (12.5)            & 44 (32.1)            &                      \\ \hline
Activities of Daily Living score   &                                     & 1.43 (1.91)          & 1.18 (1.82)          & .186                 \\ \hline
Duke Activity Status Index         &                                     & 18.86 (6.71)         & 19.73 (7.09)         & .190                 \\ \hline
Do-not-resuscitate status          &                                     & 45 (9.4)             & 3 (2.2)              & .010                 \\ \hline
Cancer status                      &                                     &                      &                      & .358                 \\
                                   & Metastatic                          & 41 (8.5)             & 8 (5.8)              &                      \\
                                   & Yes                                 & 70 (14.6)            & 16 (11.7)            &                      \\
                                   & No                                  & 369 (76.9)           & 113 (82.5)           &                      \\ \hline
SUPPORT model survival probability &                                     & 0.70 (0.15)          & 0.67 (0.17)          & .146                 \\ \hline
APACHE III score                   &                                     & 49.09 (16.29)        & 51.52 (17.18)        & .129                 \\ \hline
Glasgow coma score                 &                                     & 5.30 (16.22)         & 6.71 (17.39)         & .378                 \\ \hline
Physiological measurements         & \multicolumn{1}{l|}{}               & \multicolumn{1}{l}{} & \multicolumn{1}{l}{} & \multicolumn{1}{l}{} \\
                                   & Weight (kg)                         & 65.33 (26.23)        & 69.36 (22.47)        & .102                 \\
                                   & Temperature                         & 37.43 (1.61)         & 37.50 (1.66)         & .632                 \\
                                   & Mean blood pressure                 & 85.93 (39.47)        & 75.62 (36.08)        & .006                 \\
                                   & Respiratory rate                    & 30.48 (11.87)        & 26.39 (13.86)        & .001                 \\
                                   & Heart rate                          & 112.64 (38.45)       & 117.10 (36.39)       & .226                 \\
                                   & PaO2/FiO2 ratio                     & 256.66 (119.85)      & 225.53 (103.30)      & .006                 \\
                                   & PaCO2                               & 41.46 (14.61)        & 36.93 (10.05)        & .001                 \\
                                   & PH                                  & 7.38 (0.10)          & 7.40 (0.09)          & .182                 \\
                                   & White blood count                   & 14.54 (11.22)        & 15.83 (8.71)         & .211                 \\
                                   & Hematocrit                          & 32.42 (8.77)         & 30.81 (7.26)         & .050                 \\
                                   & Sodium                              & 135.67 (6.71)        & 135.55 (6.40)        & .848                 \\
                                   & Potassium                           & 4.07 (1.01)          & 3.89 (0.86)          & .063                 \\
                                   & Creatinine                          & 1.94 (2.14)          & 2.26 (2.23)          & .133                 \\
                                   & Bilirubin                           & 1.46 (3.38)          & 1.45 (2.17)          & .982                 \\
                                   & Albumin                             & 3.22 (0.64)          & 3.09 (0.64)          & .041                 \\ \hline
Comorbidity illness                & \multicolumn{1}{l|}{}               & \multicolumn{1}{l}{} & \multicolumn{1}{l}{} & \multicolumn{1}{l}{} \\
                                   & Cardiovascular comorbidity          & 100 (20.8)           & 36 (26.3)            & .215                 \\
                                   & Congestive heart failure            & 123 (25.6)           & 38 (27.7)            & .699                 \\
                                   & Dementia                            & 24 (5.0)             & 3 (2.2)              & .237                 \\
                                   & Psychiatric history                 & 44 (9.2)             & 6 (4.4)              & .102                 \\
                                   & Pulmonary disease                   & 124 (25.8)           & 17 (12.4)            & .001                 \\
                                   & Renal disease                       & 26 (5.4)             & 10 (7.3)             & .534                 \\
                                   & Cirrhosis, hepatic failure          & 28 (5.8)             & 9 (6.6)              & .908                 \\
                                   & Upper GI bleeding                   & 16 (3.3)             & 3 (2.2)              & .687                 \\
                                   & Tumor, leukemia, lymphoma           & 108 (22.5)           & 24 (17.5)            & .256                 \\
                                   & Immunosuppression, organ transplant & 173 (36.0)           & 50 (36.5)            & 1.000                \\ \hline
Transfer from other hospital       &                                     & 44 (9.2)             & 22 (16.1)            & .032                 \\ \hline
Definite myocardial infarction     &                                     & 22 (4.6)             & 13 (9.5)             & .048                 \\ \hline
\end{tabular}}
\end{table}